\documentclass[11pt]{article}
\usepackage{amsmath, amssymb, amscd, amsthm, amsfonts}
\usepackage{graphicx}
\usepackage{hyperref}
\usepackage[dvipsnames]{xcolor}
\usepackage{tcolorbox}
\usepackage{bm}
\usepackage[mathscr]{euscript}
\usepackage{caption}
\usepackage{subcaption}
\usepackage[normalem]{ulem}
\usepackage[style=nature, citestyle=numeric-comp, sorting=none, backend=biber, maxnames = 5]{biblatex}
\bibliography{References.bib}

\newcommand{\rd}{{\rm d}}
\newcommand{\e}{{\cal E}}
\newcommand{\s}{{\cal S}}

\newcommand{\cop}[3][-]{\hat{#2}_{#1 \bf #3}^{\dagger}}
\newcommand{\aop}[3][]{\hat{#2}_{#1 \bf #3}} 
\newcommand{\vt}[1]{{\small \bf #1}} 
\newcommand{\Tr}{{\rm Tr}}

\newcommand{\rred}{\hat{\rho}_{\rm red}}
\newcommand{\Ri}{\tilde{\rho}_{\rm red}}
\newcommand{\dg}{\dagger}

\newcommand{\h}[1]{{\hat #1}}

\begin{document}


{\renewcommand{\thefootnote}{\fnsymbol{footnote}}
		
\begin{center}
{\LARGE Time-convolutionless cosmological master equations:\\[1.5mm] Late-time resummations and decoherence for non-local kernels 
} 
\vspace{1.5em}

Suddhasattwa Brahma$^{1}$\footnote{e-mail address: {\tt suddhasattwa.brahma@gmail.com}}, 
Jaime Calder\'on-Figueroa$^{2}$\footnote{e-mail address: {\tt jrc43@sussex.ac.uk}}
and 
Xiancong Luo$^{1}$\footnote{e-mail address: {\tt edlxc.35@gmail.com}}
\\
\vspace{1.5em}

$^{1}$Higgs Centre for Theoretical Physics, School of Physics and Astronomy,\\ University of Edinburgh, Edinburgh EH9 3FD, UK\\[2mm]

$^2$Astronomy Centre, University of Sussex, Falmer, Brighton, BN1 9QH, UK\\[2mm]

\vspace{1.5em}
\end{center}
}

\setcounter{footnote}{0}

\begin{abstract}
\noindent We revisit a simple toy model of two scalar fields in de Sitter space, playing the roles of ``system'' and ``environment'' degrees of freedom, which interact with each other. We show that there are secular divergences in physically relevant observables which arise solely from the non-Markovian part of the memory kernel, contrary to popular belief that secular growth typically comes from local terms in evolution equations. Nevertheless, we show that these terms can still be non-perturbatively resummed, using the time-convolutionless master equation formalism, which improves upon previous approximations. At the same time, there are other physical quantities in the same model that are dominated by local terms in the memory kernel. Therefore, we conclude that, for cosmological backgrounds, either the dissipation or the noise kernel can end up being dominated by non-local terms depending on the nature of the system-environment coupling.
\end{abstract}

\section{Introduction}
Inflation -- the most widely accepted paradigm for the early-universe -- predicts that the large scale structure of the cosmos has its origins in the quantum vacuum fluctuations of the scalar field responsible for the background accelerated expansion \cite{starobinsky1980new, fang1980entropy, Guth:1980zm, Sato:1981qmu, Linde:1981mu,Albrecht:1982wi}. However, a deeper dive into this formalism shows that one needs to break up the same inflaton field into the homogeneous background, sourcing the evolution of the universe, and its quantum fluctuations, responsible for structure formation \cite{Mukhanov:1982nu}. A natural question that follows is what is the backreaction \cite{brandenberger2002back, PerreaultLevasseur:2014ziv} of the sub-horizon quantum fluctuations on the effective field theory (EFT) of the `classicalized' super-Hubble  modes \cite{Cohen:2020php, burgess2015eft}? Another way to phrase this problem is as follows: The equal-time in-in correlations of a massless scalar field in quasi-de Sitter (dS) space give us the spectra of the adiabatic perturbations which can be verified in the temperature fluctuations in the CMB \cite{collaboration2020planck}. Nevertheless, issues of infrared (IR) singularities and late-time secular divergences \cite{burgess2015eft, Burgess:2015ajz, Cespedes:2023aal} are know to afflict such computations when considering loop corrections \cite{senatore2010loops, Assassi:2012et}. 

These questions are typically tackled using Starobinsky's stochastic formulation of inflation\cite{Starobinsky:1979ty, Morikawa:1989xz,nambu1988stochastic,nambu1989stochastic,kandrup1989stochastic,nakao1988stochastic,mollerach1991stochastic,linde1994big,starobinsky1994equilibrium}. The general idea behind it is that gradient terms are expensive once the wavelength of a perturbation becomes super-Hubble, and thus its evolves essentially independently outside the horizon. This allows one to write down a Fokker-Planck equation for the probability distribution of the scalar field, coarse-grained over each Hubble volume. Its evolution is a balance between the classical drift due to the quasi-dS scalar potential and `quantum kicks', modeled by a Gaussian white noise, due to the sub-horizon fluctuations. Since it offers a gradient expansion for the super-Hubble density perturbation, instead of the one in terms of its amplitude, it can make predictions regarding the tail of its probability distribution where linear perturbation theory is bound to fail. And in this way, it can make nontrivial predictions for, say, the production of primordial black-holes during inflation (see, for instance, \cite{Vennin:2020kng,vennin2024quantum, Panagopoulos:2019ail}).

Stochastic inflation is supposed to also be precisely the non-perturbative formalism designed to cure the IR issues of massless fields during inflation. Since it is expected to provide the leading order effect in the resummation of late-time IR divergences in cosmology\footnote{For partial results in this direction, see \cite{Seery:2007we,Enqvist:2008kt,seery2009parton,Burgess:2009bs,Seery:2010kh,Gautier:2013aoa,Guilleux:2015pma,Gautier:2015pca,Hardwick:2017fjo,Markkanen:2017rvi,LopezNacir:2019ord,Gorbenko:2019rza,Mirbabayi:2019qtx,Adshead:2020ijf,Moreau:2020gib}.}, its assumptions need to be carefully examined. For instance, one of the primary approximations for stochastic inflation is made in the computation of the quantum noise due to the short wavelength fluctuations. Generally, it is typically assumed to be white \cite{Finelli:2008zg, Onemli:2015pma, Calzetta:1995ys, Garbrecht:2013coa, Vennin:2015hra, burgess2015eft,Baumgart:2019clc,Baumgart:2020oby}, although `colored' corrections are expected to come from non-Gaussian contributions \cite{Cohen:2021fzf, Cohen:2021jbo}. While it has been argued that both the choice of the `sharp' window function \cite{Mahbub:2022osb,Andersen:2021lii} (to demarcate the short and long wavelength scales) as well as the `test scalar' approximation  \cite{Boyanovsky:1994me} can invalidate it, nevertheless this `Markovian' assumption\footnote{At the very outset, we want to emphasize that we do \textbf{not} use Markovian and time-locality interchangeably in this draft, as is sometimes erroneously done in the cosmology literature. Also, we loosely say that a white noise implies a Markovian system in the classical sense even though we will define what we mean by Markovianity in a quantum setup as having a semi-group property later on \cite{Prudhoe:2022pte}.} of having a white noise for the integrated-out quantum degrees of freedom (dofs) remains a ubiquitous one in cosmology (\textit{e.g.,} \cite{Mirbabayi:2020vyt, Pattison:2019hef, Gong:2019yyz}) with a few notable exceptions \cite{shandera2018open, Pinol:2020cdp, Zarei:2021dpb, Figueroa:2021zah}.

What is even more, this assumption in stochastic inflation has often been used as an argument for inferring about the origin of secular terms that are to be expected when computing loop corrections during inflation. There is a rich literature on computing one-graviton loop corrections to, say, a conformally coupled scalar field propagator in dS space \cite{Glavan1, Glavan2, Boran:2014xpa} in which it has been explicitly shown that the secular terms are the ones which come from time-local terms in the evolution equation whereas the non-local terms decay and do not contribute at late-times. Similar results have also been obtained for the one-loop graviton correction to the vacuum polarization for dynamical photons in a dS background \cite{wang2015excitation}. Such secular enhancement is generally found to independent of the nonlocal terms in a variety of models \cite{Glavan3, Glavan4, Glavan5, Degueldre:2013hba, miao2006gravitons, kahya2007quantum, Prokopec:2002uw}, and is hence free of any \textit{memory} effects. These results lend support to the general expectation that stochastic inflation captures all the leading IR secular terms via a local equation, and thus the Markovian approximation is a sufficient one for cosmology. 

In this paper, we will revisit an old model \cite{Boyanovsky:2015tba} to show that indeed secular divergences can just as easily come from the non-local part of the memory kernel -- an object that captures correlations of the integrated-out  fields. More concretely, we will show explicitly that secularly divergent terms come from non-local terms in the (dissipation) part of the kernel and, consequently, that it is not necessary that late-time behaviour is always dominated by local terms. More importantly, we shall demonstrate that such secular divergences, arising from non-Markovian terms, can nevertheless be exactly resummed, without invoking any additional, \textit{ad hoc} approximations, using an appropriate master equation formalism. In our example, we shall give the physical interpretation of different co-efficients of the master equation, corresponding to different parts of the kernel, which will be dominated by local and non-local terms at late-times. The main purpose of our work is thus to show that:
\begin{enumerate}
	\item Secular divergences can just as easily stem from non-Markovian terms and, more importantly, such terms can still be resummed at late times following a precise algorithm that does not involve any arbitrary approximations. 
	\item The memory kernel, corresponding to the integrated-out (or coarse-grained) fields, in the same model can affect different physical quantities differently. More specifically, local and non-local parts of the kernel can become dominant for different physical observables.
\end{enumerate}

In the next section, we briefly discuss master equations in cosmology before setting up the one relevant for us in Sec-3. We discuss physical interpretations of coefficients of the master equations as they pertain to dissipative corrections to cosmological observables and diffusion terms leading to decoherence of the quantum fields. In Sec-4, we compute corrections to the power spectrum and the purity of the (system) density matrix and show how they are dominated by the non-local and local parts of the kernel, respectively. Finally, we conclude by summarizing our findings and outlining their applications to future work.

\section{Open quantum systems in cosmology}\label{ME_intro}
To achieve the above objectives, we shall adopt an open EFT approach \cite{calzetta2009nonequilibrium,breuer2002theory} to the problem, an idea that has been gaining considerable importance over the last decade in cosmology \cite{Burgess:2022rdo,burgess2016open,Colas:2022hlq,Colas:2022kfu,Colas:2023wxa, Colas:2024lse,Colas:2024xjy,Colas:2024ysu,Salcedo:2024smn,Boyanovsky:2015xoa,Boyanovsky:2015jen,Boyanovsky:2015tba,hollowood2017decoherence, shandera2018open, Choudhury:2018ppd,Brahma:2020zpk,Brahma:2021mng, Brahma:2022yxu, Burgess:2022nwu,banerjee2023thermalization,Kaplanek:2022xrr,Cao:2022kjn,Prudhoe:2022pte,Kading:2022hhc,Kading:2022jjl,Kading:2023mdk,Alicki:2023rfv,Alicki:2023tfz,Creminelli:2023aly,Keefe:2024cia,Bowen:2024emo,crossley2017effective, Colas:2022hlq, Burgess:2015ajz}. The main realization here is that open quantum systems are particularly suited to spacetimes with horizons, such as the one for accelerated expansion. There is a sector of the Hilbert space that is hidden from the observer, thus being dubbed ``environment'' $(\e)$, which can still exchange energy and information with the ``system'' $(\s)$ sub-sector. The usefulness of using an open quantum system is that it is able to coarse-grain the effects of the high energy or sub-Hubble dofs on the dynamics of the long-wavelength modes when there is no exclusion principle (such as the conservation of energy) available.

Hence, the master equation aims to capture the evolution of the system dofs by integrating (or tracing) out the environment modes. One of the main advantages of this formalism is that it can capture reliable late-time dynamics of the system by overcoming the \textit{secular growth} associated with time-dependent backgrounds in cosmology. The underlying reason for secular growth at late times is that gravitational interactions are universal, and once turned on, it is not possible to isolate the system from the ever-present medium, namely, gravity. This implies that even in the weakly-interacting regime, for long time-intervals small effects can accumulate to lead to the failure of perturbation theory. Resummation of such late-time divergences typically implies deriving the master equation at some order in perturbations theory, say $\lambda^2$, and then treating it as the \textit{bona fide} dynamical evolution equation that needs to be solved \textit{non-perturbatively} without performing any expansion \cite{Colas:2022hlq}. This means that the master equation would contain all term at $\mathcal{O}(\lambda^2)$ and some terms at $\mathcal{O}(\lambda^{n>2})$. 

The details of much of this discussion can be found in, say \cite{breuer2002theory,Colas:2022hlq}, and we only very briefly touch upon the time-convolutionless master equation formalism that we will employ for our model. The density matrix, corresponding to the entire system $(\s + \e)$ undergoes unitary evolution governed by the Louville-von Neumann equation:
\begin{equation}
	\frac{\rd}{\rd \tau}\rho_I (\tau) = -i \left[H_I (\tau),\rho_I (\tau)\right]\,, 
\end{equation}
where the interaction Hamiltonian has operators corresponding to both $\s$ and $\e$. For the $\s$ sub-sector, this equation can be put in an exact, closed form which determines the evolution of the reduced density matrix corresponding to $\s$, known as the Nakajima-Zwanzig master equation. However, it is an integro-differential equation that has exactly the same level of difficulty as the Louville-von Neumann equation, and thus needs to be solved using some approximation scheme.

The first of these is the so-called Born approximation, which assumes a weak coupling between $\s$ and $\e$ and expands in this interaction paarameter $\lambda$. One typically also assumes that the initial density matrix is in a factored state: $\rho(\tau_0) = \rho^\s(\tau_0) \otimes \rho^\e(\tau_0)$. Although this is not an essential assumption, it is a natural one for cosmology since we will start off both the system and environment modes in the Bunch-Davies vacuum  state. At $\mathcal{O}(\lambda^2)$, this results in a master equation of the form:
\begin{equation}\label{eq:nz2}
	\partial_t \rho^\s (\tau) = - \lambda^2 \int_{\tau_0}^\tau \rd \tau'\ \Tr_\e \big[ H_I (\tau), \left[ H_I (\tau'), \rho^\s (\tau') \otimes \rho^\e \right] \big]\;.
\end{equation}
Since the RHS of the above equation has $\rho^\s$ depending on $\tau'$, rather than on $\tau$ alone, this equation is not amenable to late-time resummations. This is because assuming that the master equation generates dynamics non-perturbatively, even though it is derived at $\mathcal{O}(\lambda^2)$, works only because it is not dependent on some specific value of time. The next popular approximation is the Markovian one where the above master equation reduces to the standard Lindblad form. However, in this work, we shall not invoke this restrictive limit which basically assumes that the master equation generates dynamics that fulfills a semi-group property, \textit{i.e.}, $\rho^\s(\tau') = \mathcal{L}_{\tau \rightarrow \tau'} \ \rho^\s(\tau)$ where $\mathcal{L}_{\tau \rightarrow \tau'} =  \mathcal{L}_{\tau \rightarrow \tau''}  \  \mathcal{L}_{\tau'' \rightarrow \tau'}$.

Instead, we will work with the time-convolutionless approximation, based on a cumulant expansion (see \cite{breuer2002theory} for details). In this case, one can rewrite the master equation, at leading order (hereafter the TCL$_2$ master equation), as 
\begin{equation}\label{eq:tcl2}
	\partial_t \rho^\s (\tau) = - \lambda^2 \int_{\tau_0}^\tau \rd \tau'\ \Tr_\e \big[ H_I (\tau), \left[ H_I (\tau'), \rho^\s (\tau) \otimes \rho^\e \right] \big]\;.
\end{equation}
Thus, at this order, the only difference is that in the TCL$_2$ master equation $\rho^\s$ is evaluated at $\tau$ instead of $\tau'$. It can be shown that the terms dropped in this equation, as compared to \eqref{eq:nz2}, is at $\mathcal{O}(\lambda^4)$ or higher. More importantly, the TCL formalism allows one to capture non-Markovian effects even though we end up working with a time-local equation. And this is exactly one of the technical issues that we wish to clarify in this work, and correct some associated errors in the literature on this topic.

Note that the reader might think that the substitution of $\rho^\s(\tau') \rightarrow \rho^\s(\tau)$ in the above master equation is nothing but an application of the Markovian approximation to the system. Indeed, this line of thinking has been mistakenly employed before \cite{Boyanovsky:2015tba, hollowood2017decoherence}. One of the goals of this work is to underscore that despite the (time-)local nature of the TCL$_2$ equation, it remains capable of adequately describing a non-Markovian system. However, a subtlety arises in the form of a residual artifact of the ``memory'' of the initial state remains in the lower limit of the integral on the RHS. In other words, when we compute the coefficients of the TCL$_2$ master equation, in terms of expressions quantifying diffusion and dissipation, we shall have to take particular care to drop such terms with explicit dependence on such initial time $\tau_0$. Such terms have been called ``spurious'' earlier in \cite{Colas:2022hlq}, and it will be shown that the TCL$_2$ cannot be resummed unless such terms are dropped by hand. Going beyond these technical aspects, the main message is that the system described by the TCL$_2$ master equation is \textit{not} a Markovian one and we will show how previous results were only approximate since it did not involve a proper analyses of such spurious terms, and how we improve upon those results in our work.

\section{A toy-model for sub- and super-Hubble mode-coupling}
Our model consists of a system comprising a minimally coupled scalar field ($\chi$) and an environment composed of a conformally coupled scalar field ($\psi$), with an interaction term which has previously been studied in \cite{Boyanovsky:2015tba, Boyanovsky:2015jen, hollowood2017decoherence}. The reason behind studying this model is two-fold. Firstly, the associated memory kernel of the environment field has the advantage of cleanly splitting into two different contributions, one that is clearly time-local while the other non-local, as we will show later on. Secondly, the conformally-coupled ``environment'' field is a good proxy for the sub-Hubble modes since the latter are least influenced by the background curvature. On the other hand, the minimally coupled  ``system'' scalar would also be assumed to be massless, thereby playing the role of the adiabatic degree of freedom. Therefore, this setup can play the role of a toy-model for stochastic inflation. Nonetheless, it is important to keep in mind that this is still a model of two interacting test scalars in a dS background: $\rd s^2 = a(\tau)^2 \left[-\rd \tau^2 + \rd {\bf x}^2\right]$ where $a(\tau)=-1/(H\tau)$ is the scalar factor, $H$ the Hubble parameter and $\tau$ denotes the conformal time. 

The action, written in terms of (canonical) Mukhanov-Sasaki fields, is given by
\begin{equation}
    {\cal S} = \int \rd\tau\ \rd^3 x\ \Big\{ \frac{1}{2} \Big[ \chi'^2 - (\nabla \chi)^2 + \frac{a''}{a} \chi^2 + \psi'^2 -(\nabla \psi)^2\Big] + \lambda  \ a  \ \chi : \psi^2 : \Big\}\;,
\end{equation}
where primes denote derivatives with respect to conformal time, $\tau$. From the above action, we can obtain the system and environment Hamiltonians, together with the interaction potential, as:
\begin{align}
    \h{H}_\s & = \frac{1}{2}\int \frac{\rd^3 k}{(2\pi)^3} \left[ \h{\pi}_{\vt k} \h{\pi}_{-\vt k} + k^2 \h{\chi}_{\vt k} \h{\chi}_{-\vt k} + \frac{a'}{a} \left( \h{\chi}_{\vt k} \h{\pi}_{-\vt k} + \h{\pi}_{\vt k} \h{\chi}_{-\vt k} \right) \right]\;, \\
    \h{H}_\e & = \frac{1}{2} \int \frac{\rd^3 k}{(2\pi)^3} \left[ \h{\pi}^{(\psi)}_{\vt k} \h{\pi}^{(\psi)}_{-\vt k} + k^2 \h{\psi}_{\vt k} \h{\psi}_{-\vt k} \right]\;, \nonumber \\
    \h{V}(\tau) & = - \lambda  a(\tau) \int \rd^3 x\ \h{\chi} (\vt{x}) : \h{\psi}^2 (\vt{x}) : \;,\label{interaction_term}
\end{align}
where $\h{\pi}$ is the momentum conjugate of $\h{\chi}$, and $\h{\pi}^{(\psi)}$ is that of $\h{\psi}$, respectively. From here on, operators in the Schrödinger picture will be denoted with hats, while interaction picture operators will be denoted with tildes. Following \cite{Boyanovsky:2015tba}, the interaction term has been normal-ordered as $:\psi^2: = \psi^2 - [{\rm Tr}(\rho_0 \psi)]^2$, where $\rho_0$ is the initial (product) density matrix. Given the free Hamiltonians, the Heisenberg equations of motion can be evaluated as
\begin{align}
    \chi''_{ k} (\tau) + \left[ k^2 - \frac{2}{\tau^2} \right]\chi_{ k} (\tau) = 0\,, \qquad \psi''_{ k} (\tau) + k^2 \psi_{ k} (\tau) = 0 \,,
\end{align}
with solutions, assuming Bunch-Davies initial conditions, given by
\begin{align}\label{eq:bd}
     \chi_{k} (\tau) = \frac{e^{-i k \tau}}{\sqrt{2k}} \left( 1 - \frac{i}{k \tau} \right)\,, \qquad  \psi_k (\tau) = \frac{e^{-i k \tau}}{\sqrt{2k}}\;,
\end{align}
and similary for their conjugate momenta:
\begin{equation}\label{eq:bdp}
    \pi_k (\tau) = \chi_k' (\tau) - \frac{a'}{a}\chi_k (\tau) = -i \sqrt{\frac{k}{2}}e^{-ik\tau}\;, \qquad \pi^{(\psi)}_k (\tau) = -i \sqrt{\frac{k}{2}} \left(1 + \frac{i}{k\tau}\right)\;.
\end{equation}

Consequently, the quantum operators for the system field (and its conjugate momenta) read
\begin{align}
    \tilde{\chi}_{\vt k} (\tau) & = \chi_k (\tau) \aop[]{a}{k} + \chi_k^* (\tau) \cop[-]{a}{k}\;, \label{eq:squant} \\
     \tilde{\pi}_{\vt k} (\tau) & = \pi_k (\tau) \aop[]{a}{k} + \pi_k^* (\tau) \cop[-]{a}{k}\;, \label{eq:pquant}
\end{align}
where $\aop[]{a}{k}$ and $\cop[-]{a}{k}$ are the annihilation and creation operators at $\tau_0$, respectively. 

Finally, let us express these fields at arbitrary times in a time-local way, as required in the TCL framework. For this, we `solve' \eqref{eq:squant} and \eqref{eq:pquant} for $\aop[]{a}{k}$ and $\cop[-]{a}{k}$ at $\tau$, and subsequently substitute these solutions back into the corresponding equations at $\tau'$, yielding
\begin{align}
    \tilde{\chi}_{\vt p} (\tau') & = 2\ {\rm Im}\left[\chi_p (\tau') \pi_p^* (\tau)\right] \tilde{\chi}_{\vt p} (\tau) - 2\ {\rm Im}\left[\chi_p (\tau') \chi_p^*(\tau)\right] \tilde{\pi}_{\vt p} (\tau)\,, \label{eq::ttr1}\\
    \tilde{\pi}_{\vt p} (\tau') & = 2\ {\rm Im}\left[ \pi_p^*(\tau) \pi_p (\tau') \right] \tilde{\chi}_{\vt p} (\tau) - 2\ {\rm Im}\left[ \chi_p^* (\tau) \pi_p (\tau') \right] \tilde{\pi}_{\vt p} (\tau)\;. \label{eq::ttr2}
\end{align}
Plugging the mode functions above, one obtains
\begin{align}
    \tilde{\chi}_{\vt p} (\tau') & = \left[ \cos (p(\tau - \tau')) + \frac{\sin(p(\tau-\tau'))}{p\tau'} \right]\tilde{\chi}_{\vt p} (\tau) \nonumber \\
    & - \left[ \frac{\tau' - \tau}{p^2 \tau \tau'} \cos(p(\tau-\tau')) + \frac{1}{p}\left(1 + \frac{1}{p^2 \tau\tau'} \right) \sin(p(\tau-\tau')) \right] \tilde{\pi}_{\vt p} (\tau)\;, \\
    \tilde{\pi}_{\vt p} (\tau') & = p \sin(p(\tau-\tau')) \tilde{\chi}_{\vt p} (\tau) + \left[ \cos(p(\tau-\tau')) - \frac{1}{p\tau} \sin(p(\tau - \tau')) \right] \tilde{\pi}_{\vt p} (\tau)\;.
\end{align}

\subsection{Building the TCL$_2$ master equation}
Following the recipe outlined in Sec-2, some of the details of which can be found in \cite{Colas:2022hlq, us:QSR}, we can write down the TCL$_2$ master equation as\footnote{We have used $\rho^\s$ and $\rho_{\rm red}$ interchangably in this paper.}
\begin{align}\label{eq:me0}
    \frac{\rd \Ri}{\rd \tau} = & - \sum_{\vt p} \lambda a(\tau) \int_{\tau_0}^{\tau} \rd \tau'\ \lambda a(\tau') \bigg\{ \left[ \tilde{\chi}_{\vt p} (\tau) \tilde{\chi}_{-\vt p} (\tau') \Ri (\tau) - \tilde{\chi}_{-\vt p} (\tau') \Ri (\tau) \tilde{\chi}_{\vt p} (\tau) \right]K_{\vt p} (\tau,\tau') \nonumber \\
    & - \left[ \tilde{\chi}_{\vt p} (\tau) \Ri(\tau) \tilde{\chi}_{-\vt p} (\tau') - \Ri (\tau) \tilde{\chi}_{-\vt p} (\tau') \tilde{\chi}_{\vt p} (\tau) \right] K_{\vt p}^* (\tau, \tau') \bigg\}\;,
\end{align}
where the memory kernel is given by\footnote{To write the solution of the integral in the quoted form, one invokes the Sokhotski–Plemelj theorem, which states 
\begin{equation*}
    \lim_{\epsilon \rightarrow 0^+} \frac{1}{x \pm i \epsilon} = {\cal P} \left( \frac{1}{x} \right) \mp i \pi \delta(x)\;,
\end{equation*}
where ${\cal P}$ denotes the principal part which shall be expressed more explicitly later on.}
\begin{align}
    K_{\vt p}(\tau,\tau') & = 2 \int \frac{\rd^3 k}{(2\pi)^3}\ \psi_k (\tau) \psi_k^* (\tau') \psi_q (\tau) \psi_q^* (\tau')\;, \quad p = |\vt{k} + \vt{q}| \nonumber \\
    & = - \frac{i}{8\pi^2} e^{-i p (\tau-\tau')} {\cal P}\left( \frac{1}{\tau - \tau'}\right) + \frac{1}{8\pi} \delta(\tau - \tau')\;. \label{eq:kr}
\end{align}
Given the system field appears at most at quadratic order in the the full Lagrangian, the evolution equation for the density matrix can be written as a sum over independent momentum modes ${\bf p}$ without any mode-coupling. This is another technical advantage of choosing the type of coupling that is being considered in this model (as opposed to, say, a $\chi^2 \psi^2$ coupling term).

It will be more compact to express the field and its conjugate momentum as coordinates of phase space, denoted as $\tilde{z}_1 (\tau) = \tilde{\chi}_{\vt p} (\tau)$, $\tilde{z}_2 (\tau) = \tilde{\pi}_{\vt p} (\tau)$. Using \eqref{eq::ttr1},\eqref{eq::ttr2}, the resulting master equation can be written as 
\begin{align}
    \frac{\rd \Ri}{\rd \tau} & = - \frac{1}{2} \sum_{\vt p} \Big\{ [D_{11} + i \Delta_{11}] (\tilde{z}_1 \tilde{z}_1^\dg \Ri - \tilde{z}_1^\dg \Ri \tilde{z}_1) + 2 [D_{12} + i \Delta_{12}] (\tilde{z}_1 \tilde{z}_2^\dg \Ri - \tilde{z}_2^\dg \Ri \tilde{z}_1) \nonumber \\
    & - [D_{11} - i \Delta_{11}](\tilde{z}_1 \Ri \tilde{z}_1^\dg - \Ri \tilde{z}_1^\dg \tilde{z}_1) - 2 [D_{12} - i \Delta_{12}](\tilde{z}_1 \Ri \tilde{z}_2^\dg - \Ri \tilde{z}_2^\dg \tilde{z}_1) \Big\}\;,
\end{align}
where all the operators are evaluated at $\tau$, and the functions are given by the following integrals: 
\begin{align}
    D_{11} & = 4 \lambda a(\tau) \int_{\tau_0}^{\tau} \rd \tau' \lambda(\tau')\ {\rm Im}\left[\chi_p (\tau') \pi_p^* (\tau)\right] {\rm Re}\left[K_p (\tau,\tau')\right]\;, \label{d11i}\\
    \Delta_{11} & = 4 \lambda a(\tau) \int_{\tau_0}^{\tau} \rd \tau' \lambda a(\tau')\ {\rm Im}\left[\chi_p (\tau') \pi_p^* (\tau)\right] {\rm Im}\left[K_p (\tau,\tau')\right]\;, \label{dt11i}\\
    D_{12} & = -2 \lambda a(\tau) \int_{\tau_0}^{\tau} \rd \tau' \lambda a(\tau')\ {\rm Im}\left[\chi_p (\tau') \chi_p^*(\tau)\right] {\rm Re}\left[ K_p (\tau,\tau') \right]\;, \label{d12i}\\
    \Delta_{12} & = -2 \lambda a(\tau) \int_{\tau_0}^{\tau} \rd \tau' \lambda a(\tau')\ {\rm Im}\left[\chi_p (\tau') \chi_p^*(\tau)\right] {\rm Im}\left[ K_p (\tau,\tau') \right]\;. \label{dt12i}
\end{align}
We leave the solutions of these integrals for the next section. 

At this point, let us consider the lower limit of the (formal) antiderivative of the above functions. Following the convention adopted in \cite{Colas:2022hlq}, we can rewrite the RHS of each of the above equations as $F_i(\tau, \tau) - F_i(\tau, \tau_0)$, the abstract index $i$ denoting the different functions for the four different quantities above. The second term, coming from the lower limit of the integral, has the memory of the initial state and is the one that has been termed `spurious' in  \cite{Colas:2022hlq}. In the Appendix, we compute these terms explicitly and show that they dominate at late-times near the conformal boundary. What this implies is that keeping such terms would not allow for the TCL$_2$ master equation to be resummed. Moreover, as has also been noticed in \cite{Colas:2023wxa}, such terms are absent in a perturbative treatment. In other words, if one were to solve the master equation by expanding the density matrix to a fixed order in $\lambda^n$, then these terms get cancelled amongst themselves. However, when solving the TCL$_2$ equation non-perturbatively by considering it as the authentic dynamical generator, we only keep some of the terms at orders higher than $\lambda^2$, and therefore such spurious terms need to be removed by hand. However, getting rid of these spurious terms is not the same as imposing a Markovian approximation. As will be shown later on, some of the master equation coefficients in \eqref{d11i}-\eqref{dt12i} have contributions \textit{solely} from the non-local part of the memory kernel and hence, by definition, has a non-Markovian origin. It is important to point out that previous papers where this model has been studied lacks any discussion of these spurious terms and we shall explain later on in the next section why their results are, therefore, necessarily approximate when considering non-perturbative resummations.

With these expressions at hand, the master equation is written in a rather economical and suggestive way, 
 \begin{align}
    \frac{\rd \Ri}{\rd \tau} & = \sum_{\vt p}\bigg( -\frac{i}{2} \Delta_{ij}(\tau) \big[\tilde{z}_i (\tau) \tilde{z}^\dg_j (\tau), \Ri (\tau) \big] - \frac{1}{2} D_{ij} (\tau) \left[ \tilde{z}_i (\tau), \big[ \tilde{z}^\dg_j (\tau), \Ri(\tau) \big]\right] \nonumber \\
    & - \frac{i}{2} \Delta_{12} (\tau) \omega_{ij} \left[ \tilde{z}_i (\tau), \big\{ \tilde{z}^\dg_j (\tau), \Ri(\tau) \big\} \right] \bigg)\;,
\end{align}
where $\omega_{ij}$ is the antisymmetric matrix
\begin{equation}
 \omega_{ij} = \begin{pmatrix}
0 & 1 \\
-1 & 0 
\end{pmatrix}\;,
\end{equation}
and $D_{ij}$ and $\Delta_{ij}$ are symmetric matrices, whose entries are given by \eqref{d11i}-\eqref{dt12i}. Noticeably, there is no $D_{22}$ term present, a consequence of the specific form of the interaction involving only configuration field-field coupling. On the other hand, in interactions involving time derivatives of the fields, one anticipates the emergence of such terms (as that would involve a field-momentum coupling  \cite{Colas:2022kfu}). The master equation, presented in this manner, bears resemblance to stochastic equation describing (quantum) Brownian motion, such as the one obtained from the Caldeira-Leggett model \cite{Colas:2022hlq, Bhattacharyya:2024duw}.

For practical purposes, it is more useful to work in the Schr\"odinger picture where the master equation takes the form:
\begin{equation}\label{sme}
    \frac{\rd \rred}{\rd \tau} = \sum_{\vt p}\bigg( -i H_{ij}^{(2)}\left[ \h{z}_i \h{z}_j^\dg\,, \rred (\tau) \right] - \frac{1}{2} D_{ij} (\tau) \left[ \h{z}_i\,, \big[ \h{z}_j^\dg\,, \rred (\tau) \big]\right]  - \frac{i}{2}\Delta_{12} \omega_{ij} \left[ \h{z}_i\,, \big\{ \h{z}_j^\dg\,, \rred (\tau) \big\} \right] \bigg)\;,
\end{equation}
with the effective quadratic Hamiltonian given by
\begin{equation}
    H_{ij}^{(2)} = \frac{1}{2} \left[ \h{z}_2 \h{z}^\dg_2 + (k^2 + \Delta_{11}) \h{z}_1 \h{z}^\dg_1 + \left(\frac{a'}{a} + \Delta_{12} \right) \left( \h{z}_1 \h{z}_2^\dg + \h{z}_2 \h{z}_1^\dg \right)  \right]\;.\label{Eff_Ham}
\end{equation}
Finally, we can rewrite the master equation in a Lindblad--like form as
\begin{align}
    \frac{\rd \rred}{\rd \tau} =& \sum_{\vt p}\bigg( -i H_{ij}^{(2)} \big[ \h{z}_i \h{z}_j^\dg \,, \rred (\tau) \big] + \gamma_{ij} (\tau) \Big( \h{z}_i \rred (\tau) \h{z}_j^\dg - \frac{1}{2} \{ \h{z}_j^\dg \h{z}_i\,, \rred (\tau) \} \Big) \bigg)\;,
\end{align}
where $\gamma_{ij} \equiv D_{ij} - i\Delta_{12} \ \omega_{ij}$ is known as the dissipator matrix.

\subsection{The dissipator matrix: Dissipation and Diffusion}
Let us focus on the coefficients of the master equation, some of which appear in the effective Hamiltonian and the others as entries of the dissipator matrix. Before presenting the explicit form of these expressions, let us delve into the physical implications of these terms. The real and imaginary part of the memory kernel $K_p(\tau, \tau')$ correspond to the noise and the dissipation kernel, respectively\footnote{For thermal states, the real and imaginary parts are related to each other via the \textit{fluctuation-dissipation} theorem. Such dynamics takes place in scenarios like warm inflation \cite{Berera:1995wh,Berera:1995ie}.}. In terms of the RHS of the dissipator matrix written above, $D_{ij}$ corresponds to the noise kernel while $\Delta_{12}$ the dissipation kernel, respectively. Both these terms signify non-unitary contributions to the reduced density matrix due to interactions with the environment field. On the other hand, $\Delta_{11}$ in \eqref{Eff_Ham} corresponds to a unitary term renormalizing the energy spectrum of the system Hamiltonian. Since the $D_{ij}$ term is responsible for diffusion, it leads to decoherence of the quantum modes associated with the system dofs. On the other hand, the $\Delta_{12}$ is the dissipative or friction term that controls the amount of energy transfer between the system and the environment and would be reflected in the equal-time correlation functions, \textit{e.g.,} the configuration field power-spectrum. ($\Delta_{12}$ plays the dual role of renormalizing the comoving Hubble parameter.) The utility of our model is that it would be clear which of these quantities are dominated by the local and the non-local parts of the kernel, and thereby which physical observables would reflect the non-Markovianity of the model.

\begin{figure}[h!]
     \centering
     \begin{subfigure}[b]{0.48\textwidth}
         \centering
         \includegraphics[width=1.1\textwidth]{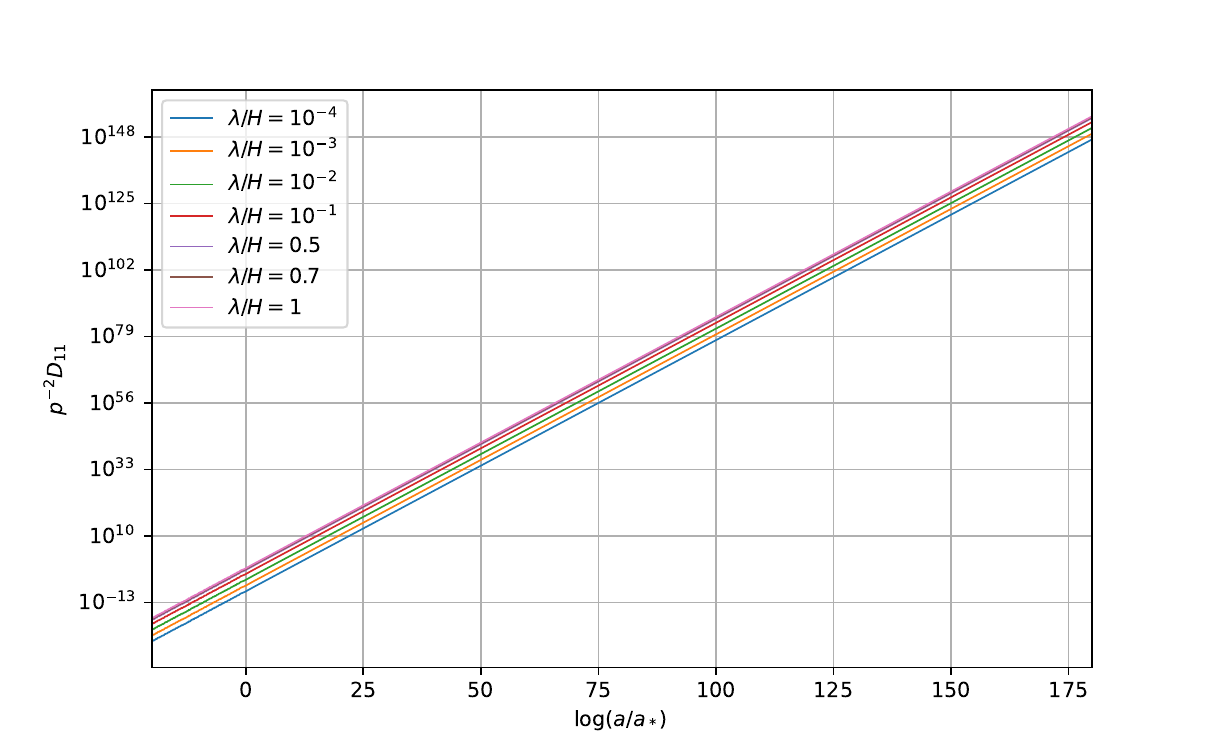}
         \caption{}
     \end{subfigure}
     \hfill
     \begin{subfigure}[b]{0.48\textwidth}
         \centering
         \includegraphics[width=1.1\textwidth]{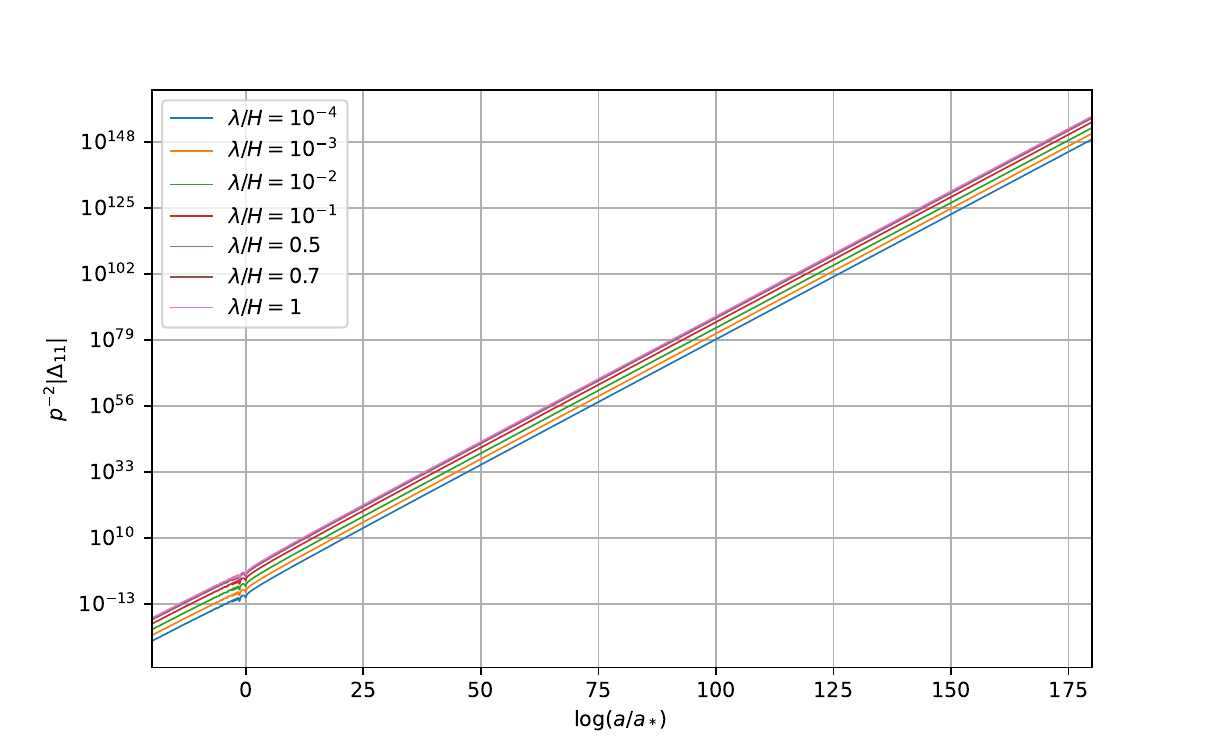}
         \caption{}
     \end{subfigure}
        \caption{$p^{-2}D_{11}$ (left) and $p^{-2}|\Delta_{11}|$ (right) for various $\lambda/H$ ratios. For (b), pre-horizon crossing, the function oscillates between positive and negative values, eventually stabilizing at the positive values depicted in the plot post-horizon crossing.}
        \label{fig:11}
\end{figure}

\begin{figure}[h!]
     \centering
     \begin{subfigure}[b]{0.48\textwidth}
         \centering
         \includegraphics[width=1.1\textwidth]{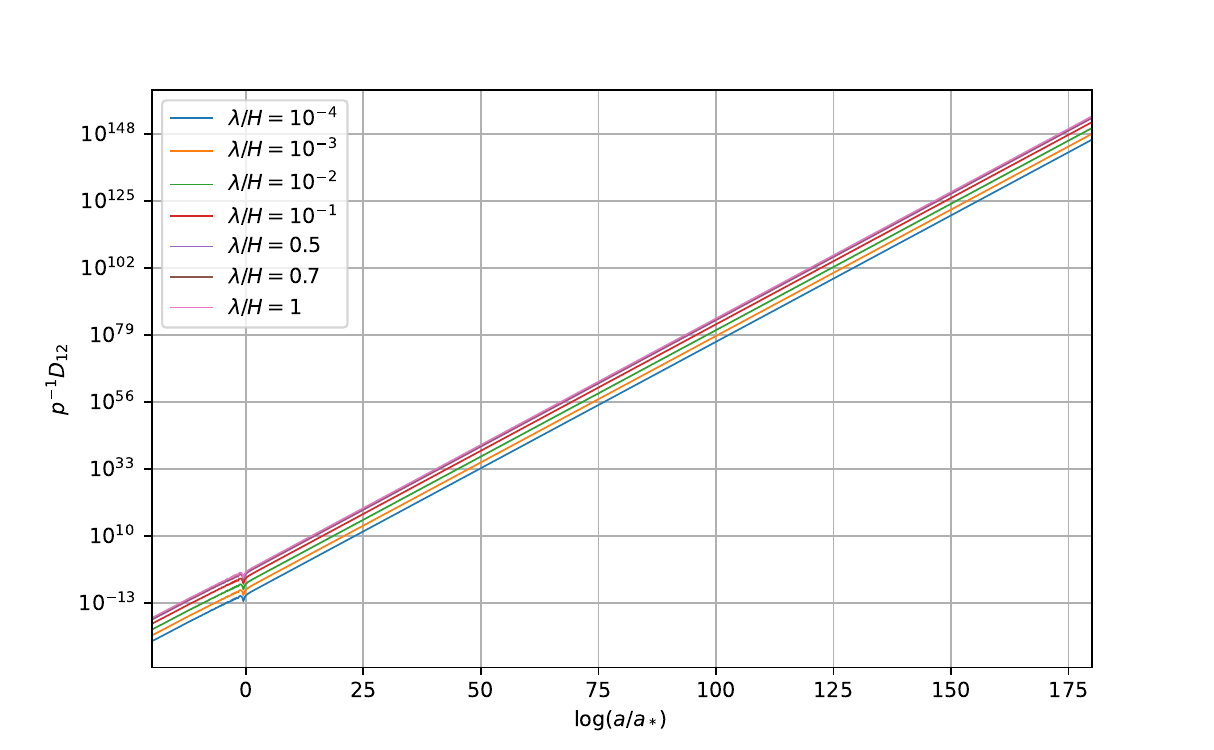}
         \caption{}
     \end{subfigure}
     \hfill
     \begin{subfigure}[b]{0.48\textwidth}
         \centering
         \includegraphics[width=1.1\textwidth]{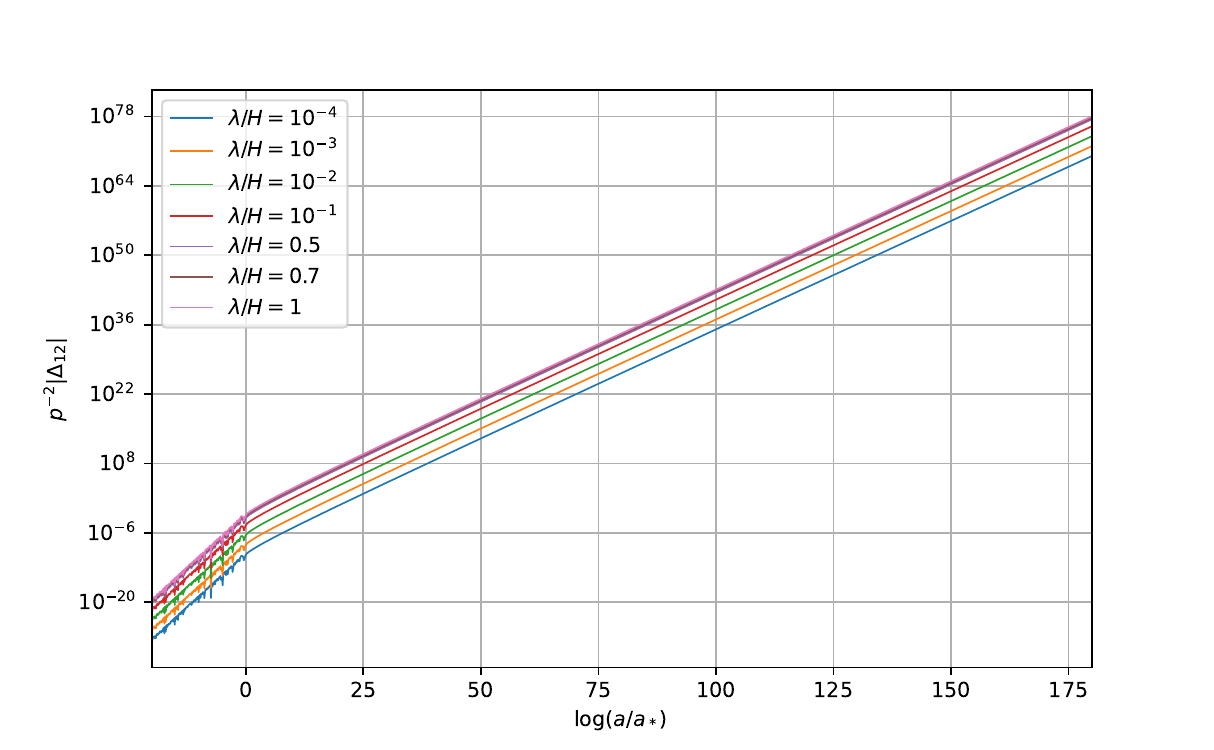}
         \caption{}
     \end{subfigure}
        \caption{$p^{-1}D_{12}$ (left) and $p^{-1}|\Delta_{12}|$ (right) for various $\lambda/H$ ratios. For (b), pre-horizon crossing, the function oscillates between positive and negative values, eventually stabilizing at the positive values depicted in the plot post-horizon crossing.}
        \label{fig:12}
\end{figure}

Keeping in mind the physical significance of $\Delta_{11}$, $D_{ij}$ and $\Delta_{12}$, we wish to find their explicit form using the integrals in \eqref{d11i}-\eqref{dt12i}. Although intricate, these integrals can indeed be evaluated analytically for this model. Particularly delicate are the terms arising from the principal value in the kernel (see \eqref{eq:kr}), \textit{i.e.,} from the time non-local term. One strategy to address this, and to isolate the associated UV-divergence, is to express the principal value as:
\begin{equation}
	{\cal P} \left( \frac{1}{\tau-\tau'} \right) = \frac{\tau-\tau'}{(\tau-\tau')^2 + \epsilon^2}\;,
\end{equation}
and subsequently expand around $\epsilon = 0^+$. We present the result of these integrals derived through this method, while relegating the expressions of the spurious $F_i (\tau,\tau_0)$ functions, originating from the lower limits of the integrals, to the  Appendix. In the following, we have highlighted in blue those terms arising from the local part of the kernel while the rest all originate from the non-local bit. Furthermore, the UV-divergent piece (also coming from the non-local part of the kernel) has been marked in red. The resulting expressions are:
\begin{align}
    D_{11} & = -\frac{\lambda^2}{8\pi^2 H^2 p \tau^3} \Big[ \gamma_E + \ln(-2p\tau) - {\rm Ci}(-2p\tau) \left[ \cos(2p\tau) + p \tau \sin(2p\tau) \right] \nonumber \\
    & + {\rm Si} (2p\tau) \left[ p\tau \cos(2p\tau) - \sin(2p\tau) \right] \Big] + {\color{blue}\frac{\lambda^2}{4\pi H^2 \tau^2}} + F_1[\tau,\tau_0]\;,\label{D11}\\
    \Delta_{11} & = \frac{\lambda^2}{8\pi^2 H^2 p \tau^3} \Big[ p\tau \left[ \gamma_E -\ln(-2 p\tau) \right] + {\rm Ci}(-2p\tau) \left[p\tau \cos(2p\tau) - \sin(2p\tau) \right] \nonumber \\
    & + {\rm Si}(2p\tau) \left[ \cos(2p\tau) + p \tau \sin(2p\tau) \right]
 \Big] + \frac{\lambda^2}{4\pi^2 H^2 \tau^2} {\color{red}\ln(2p\epsilon)} + F_2 [\tau,\tau_0]\;, \label{eq:DT11}\\
    D_{12} & = \frac{\lambda^2}{16 \pi^2 H^2 p^3 \tau^4} \Big[ (1+p^2 \tau^2) \left[ \gamma_E + \ln(-2p\tau) \right] + {\rm Ci}(-2p\tau) \left[ (-1 + p^2 \tau^2) \cos(2p\tau) - 2p\tau \sin(2p\tau) \right] \nonumber \\
    & + {\rm Si}(2p\tau) \left[ 2p\tau \cos(2p\tau) + (-1 + p^2 \tau^2) \sin(2p\tau) \right]\Big] + \textcolor{blue}{0} + F_3[\tau,\tau_0]\;,\label{D12}\\
    \Delta_{12} & = \frac{\lambda^2}{16\pi^2 H^2 p^3 \tau^4} \Big[ 2 p \tau + {\rm Ci}(-2p\tau) \left[ \sin(2p\tau) - p\tau (2 \cos(2p\tau) + p\tau \sin(2p\tau)) \right] \nonumber \\
    & + {\rm Si}(2p\tau) \left[ (-1+p^2 \tau^2) \cos(2p\tau) - 2p\tau \sin(2p\tau) \right] \Big] + F_4 [\tau,\tau_0]\;.\label{DT12}
\end{align}
We show the evolution of these functions in Figs.~\ref{fig:11} and \ref{fig:12}. The depicted plots exclusively come from the finite terms where, as promised, we have simply dropped the spurious contributions $F_i(\tau, \tau_0)$ for now. Note that we have appropriately rescaled these quantities with the correct factors of $p$ to make them dimensionless. For generating the plots and performing subsequent numerical analyses, we have selected values of $p$ such that the system evolves for 20 $e-$foldings before horizon crossing. We verified that extending this time does not alter our results. $a_*$ denotes the value of the scale-factor when the mode crosses the horizon; thus, the zero of the horizontal axis denotes horizon-crossing in the following plots.

\begin{figure}[h!]
    \centering
    \includegraphics[width=0.9\textwidth]{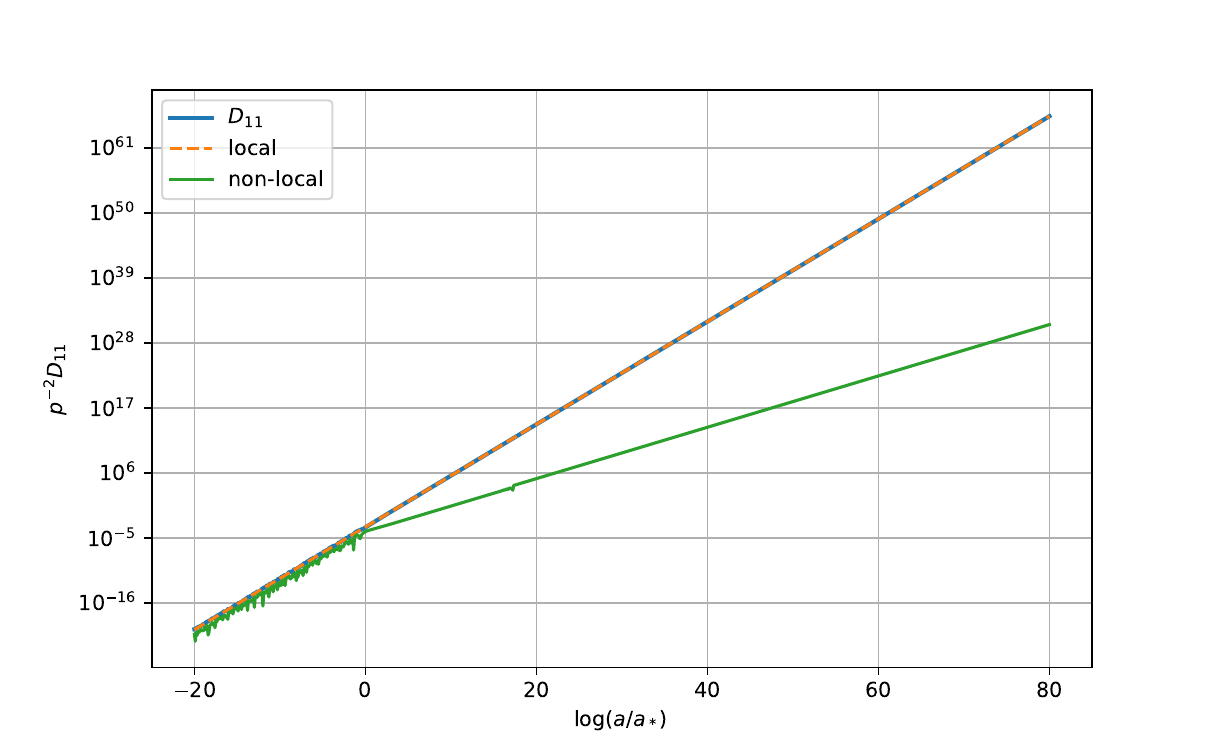}
    \caption{$D_{11}$ for a ratio $\lambda/H = 0.1$. The time-local and non-local contributions to $D_{11}$ (blue line) are plotted separately, illustrating the prevalence of the former over the course of evolution. Additionally, as the non-local contribution is negative post horizon crossing, we display its absolute value.}
    \label{fig:lnl}
\end{figure}

Let us first take a close look at the expression for $D_{11}$ as given in \eqref{D11}. The term coming from the local part of the kernel apparently looks subdominant ($\propto 1/\tau^2$) to the nonlocal part (with a piece going as $1/\tau^{3}$). However, on tracking the evolution of these terms separately, it is clearly seen that the time-local contribution consistently dominates, becoming more pronounced over timescales of a few $e-$foldings after horizon exit (see Fig.~\ref{fig:lnl}). One can show that the Taylor expansion of $D_{11}$ around $\tau=0$ is:
\begin{equation}\label{localvnl}
    D_{11} \approx {\color{blue}\frac{\lambda^2}{4\pi H^2 \tau^2}} +\frac{\lambda^2 p}{8\pi^2 H^2  \tau}+ \mathcal{O}(\tau)
\end{equation}
What this indicates is that one of the diffusion terms ($D_{11}$) is dominated by the contribution coming from the local part of the memory kernel (at all times and  certainly after horizon-crossing). This will be important for our discussion about decoherence later on. On the other hand, $D_{12}$ only has contributions from the non-local part of the kernel since the term coming from the local bit turns out to be zero for our model. Next let us look at the expression of $\Delta_{11}$ and $\Delta_{12}$ from \eqref{eq:DT11} and \eqref{DT12}, respectively. These have contributions coming solely from the non-local part of the memory kernel which shows that the dissipation in this model is \textit{necessarily} non-Markovian. However, $\Delta_{11}$ does contain a UV-divergent term that needs to be taken care of by invoking a mass counter-term. Note that this implies that this mass term has to counteract long-time correlations such that they get screened. Or, conversely, we might say that the non-local origin of this UV-divergent term means that the time non-local history dependence of the correlations in the environment induce a mass term for the system. $\Delta_{12}$ does not have any such UV-divergent piece. Finally, note that both $\Delta_{11}$ and $\Delta_{12}$ also display secular divergence even though these now have a non-local origin.

\section{Cosmological Observables}
The purpose of tracking the time evolution of the (reduced) density matrix through the master equation is to compute observables, particularly equal-time $n$-point correlation functions. In this work, we primarily focus on 2-point correlators, where every relevant combination is encapsulated by the following matrix: 
\begin{equation}
    [\Xi_{ab}]_{\vt p} = \frac{1}{2} \left( \hat{z}_a \hat{z}_b^\dagger + \hat{z}_b \hat{z}_a^\dagger \right)\;,
\end{equation}
where each term on the RHS is to be understood as the ${\vt p}$-mode of their respective field. Observables are identified as the expectation values of these operators, represented by $\Sigma_{ab} := \langle \Xi_{ab} \rangle$. This is known as the \textit{covariance matrix}, and the differential equations governing the evolution of its entries are dubbed \textit{transport equations}. 

\subsection{The transport equations}
To derive the transport equations, let us begin by considering the expectation value of an arbitrary operator ${\hat O}$, which in the Schr\"odinger picture is given by 
\begin{equation}
    \langle {\hat O} \rangle (\tau) = \Tr_\s[ {\hat O} \rred (\tau)]\;.
\end{equation}
Then, the evolution equation of the expectation value can be found through
\begin{equation}\label{eq:ev0}
    \frac{\rd}{\rd \tau} \langle {\hat O} \rangle (\tau) = \Tr_\s \left\{ {\hat O} {\cal L}_{_{\cal P}} [\rred (\tau)] \right\}\;.
\end{equation}

Using the TCL$_2$ master equation in any of its forms, say \eqref{sme}, and leveraging the cyclic property of the trace, we can cast \eqref{eq:ev0} as a combination of expectation values (underscoring the importance of a time convolutionless master equation). In doing so, and repeatedly using 
\begin{align}
    [\h{z}_i, \h{O}] =  [\h{z}_i, \Xi_{ab}] & = \frac{1}{2} [\h{z}_i, \hat{z}_a \hat{z}_b^\dagger + \hat{z}_b \hat{z}_a^\dagger] \nonumber \\
    &= \frac{1}{2} \left\{ \h{z}_a [\h{z}_i, \h{z}_b^\dg] + [\h{z}_i, \h{z}_a] \h{z}_b^\dg \right\} + (a \leftrightarrow b) \nonumber \\
    & = \frac{1}{2} \left\{ \h{z}_a (i \omega_{ib} \delta_{\vt{k},\vt{p}} ) + (i \omega_{ia} \delta_{\vt{k}, -\vt{p}}) \h{z}_b^\dg \right\} + (a \leftrightarrow b)\;,
\end{align}
where $\h{z}_i \equiv [\h{z}_i]_{\vt k}$, we obtain
\begin{equation}
    \frac{\rd}{\rd \tau} \bm{\Sigma}_{\vt p} = \bm{\omega} \bm{H}^{(2)} (\bm{\Sigma}_{\vt p} + \bm{\Sigma}_{-\vt{p}}) - (\bm{\Sigma}_{\vt p} + \bm{\Sigma}_{-\vt{p}}) \bm{H}^{(2)} \bm{\omega} -\bm{\omega} \bm{D} \bm{\omega} - \Delta_{12} \left( \bm{\Sigma}_{\vt{p}} + \bm{\Sigma}_{-\vt{p}}  \right)\;.
\end{equation}
The first two terms on the RHS come from the unitary evolution, while the latter two are associated with diffusion and dissipation, respectively, as explained in the previous subsection. For parity conserving Hamiltonians, one can safely assume $\bm{\Sigma}_{\vt p} = \bm{\Sigma}_{-\vt p}$, yielding the simple set of transport equations
\begin{equation}\label{transporteqn}
    \frac{\rd}{\rd \tau} \bm{\Sigma}_{\vt p} = 2 \bm{\omega} \bm{H}^{(2)} \bm{\Sigma}_{\vt p} - 2 \bm{\Sigma}_{\vt p} \bm{H}^{(2)} \bm{\omega} -\bm{\omega} \bm{D} \bm{\omega} - 2 \Delta_{12} \bm{\Sigma}_{\vt{p}}\;,
\end{equation}
or, in matrix form, 
\begin{equation}\label{eq:tem}
    \begin{pmatrix}
        \Sigma_{11}' &\Sigma_{12}' \\\Sigma_{12}' &\Sigma_{22}'
    \end{pmatrix} = \begin{pmatrix}
2(a'/a)\Sigma_{11} + 2 \Sigma_{12} & \Sigma_{22} - (p^2 + \Delta_{11}) \Sigma_{11} - D_{12} - 2 \Delta_{12} \Sigma_{12} \\
\Sigma_{22} - (p^2 + \Delta_{11}) \Sigma_{11} - D_{12} - 2 \Delta_{12} \Sigma_{12} & -2 (p^2 + \Delta_{11}) \Sigma_{12} - 2 (a'/a + 2 \Delta_{12}) \Sigma_{22} + D_{11}
\end{pmatrix}\,,
  \end{equation}
where it is understood that all of the terms are computed at a specific momentum mode ${\bf p}$.


\subsubsection{A simple consistency check}
Before diving into solving the transport equations -- something that can only be achieved through numerical simulations -- let us verify that the equations yield sensible expressions for the free theory, where analytical exact results are readily available. For such a scenario, we know the elements of the covariance matrix, which are merely combinations of the mode functions described in \eqref{eq:bd} and \eqref{eq:bdp}, and their complex conjugates. Upon calculation, we find:
\begin{align}\label{eq:coru}
      \Sigma^{\rm free}_{11} (\tau) = \frac{1}{2p} \left( 1 + \frac{1}{(p\tau)^2} \right) \quad & \implies \quad \frac{\rd}{\rd \tau} \Sigma^{\rm free}_{11} (\tau) = - \frac{1}{(p\tau)^3} \nonumber \\
      \Sigma^{\rm free}_{12} (\tau) = \frac{1}{2p\tau} \quad & \implies \quad \frac{\rd}{\rd \tau} \Sigma^{\rm free}_{12} (\tau) = - \frac{1}{2p\tau^2} \nonumber \\
      \Sigma_{22}^{\rm free} (\tau) = \frac{p}{2} \quad & \implies \quad \frac{\rd}{\rd \tau} \Sigma^{\rm free}_{22} (\tau) = 0\;,
  \end{align}
which perfectly matches with the free-theory transport equations:
\begin{equation}
      \frac{\rd}{\rd \tau} \bm{\Sigma}^{\rm free} = \begin{pmatrix}
2(a'/a)\Sigma_{11}^{\rm free} + 2 \Sigma_{12}^{\rm free} & \Sigma_{22}^{\rm free} - p^2 \Sigma_{11}^{\rm free} \\
\Sigma_{22}^{\rm free} - p^2 \Sigma_{11}^{\rm free} & -2 p^2 \Sigma_{12}^{\rm free} - 2 (a'/a) \Sigma_{22} ^{\rm free}
\end{pmatrix}\;.
  \end{equation}
On a cursory glance, the transport equation for $\Sigma_{11}$, namely the configuration field ($\chi$) power spectrum, looks remarkably similar in structure to the analogous equation in \eqref{eq:tem}. However, this resemblance does not imply that interactions have no impact on the power spectrum. Quite the contrary, the influence of interactions is encapsulated in the coupling of $\Sigma_{11}$ to other equations, where the dissipative and diffusion coefficients appear explicitly. 


\subsection{Numerical Solutions for the covariance matrix}
As previously mentioned, the system of transport equations \eqref{eq:tem} requires numerical techniques to obtain a solution. Alternative approaches, like the one implemented in \cite{Boyanovsky:2015tba, Brahma:2021mng}, involve approximations in the super-Hubble regime by ignoring the contribution of the so-called decaying modes to the system. As a by-product, we shall show that these approximations render results in excellent agreement with the exact (numerical) results obtained here. 

\begin{figure}[h!]
    \centering
    \includegraphics[width=0.9\textwidth]{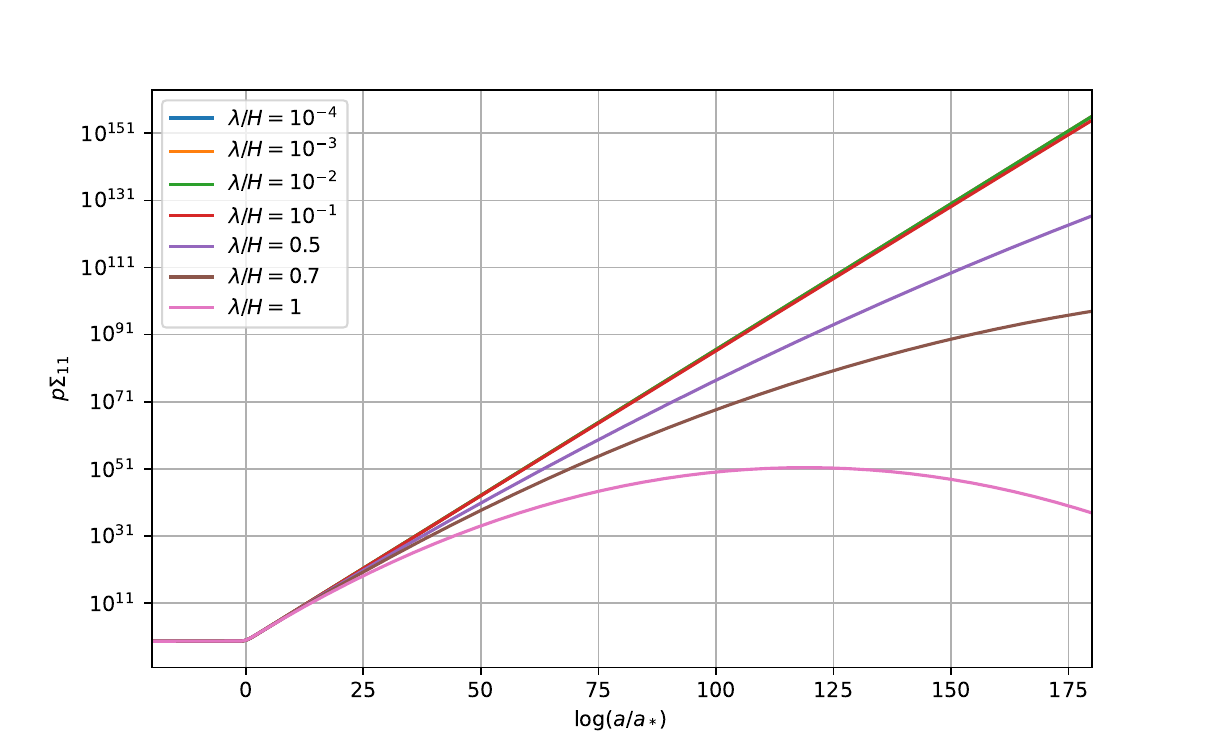}
    \caption{Numerical solution of $p\Sigma_{11}$ for different values of $\lambda/H$. }
    \label{fig:S11}
\end{figure}

\begin{figure}[h!]
    \centering
    \includegraphics[width=0.9\textwidth]{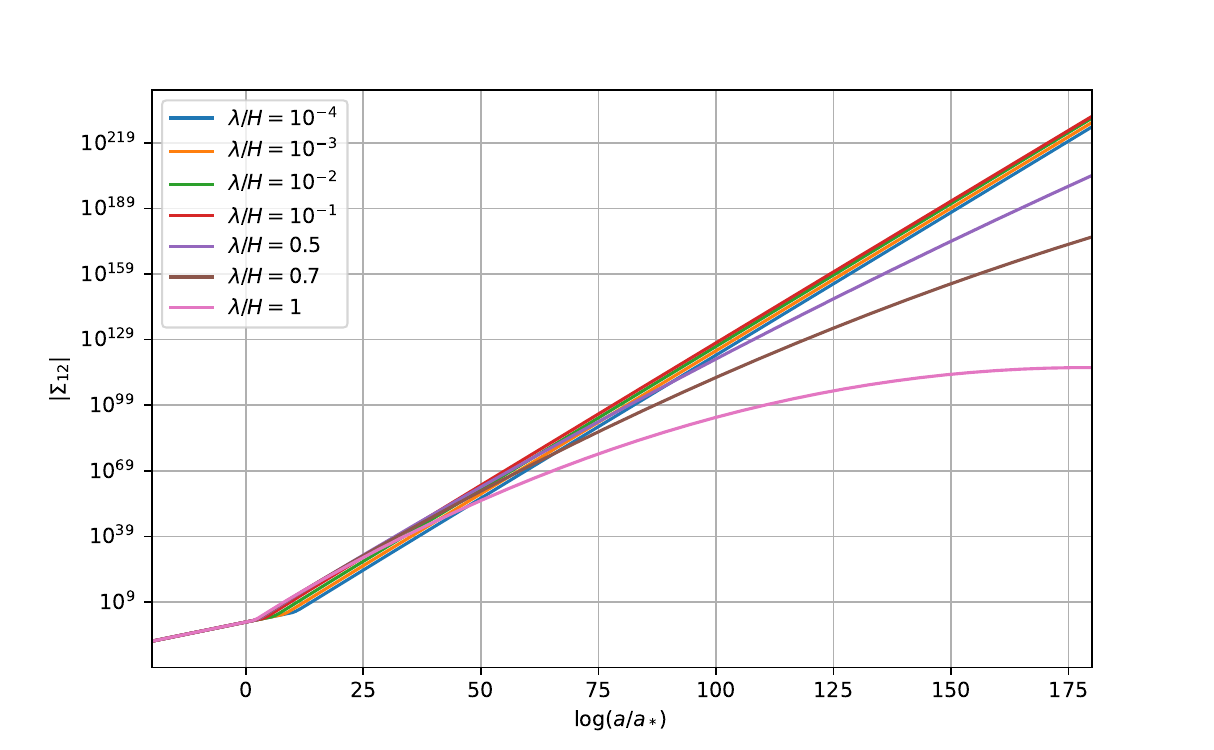}
    \caption{Numerical solution of $\Sigma_{12}$ for different values of $\lambda/H$. }
    \label{fig:S12}
\end{figure}

\begin{figure}[h!]
    \centering
    \includegraphics[width=0.9\textwidth]{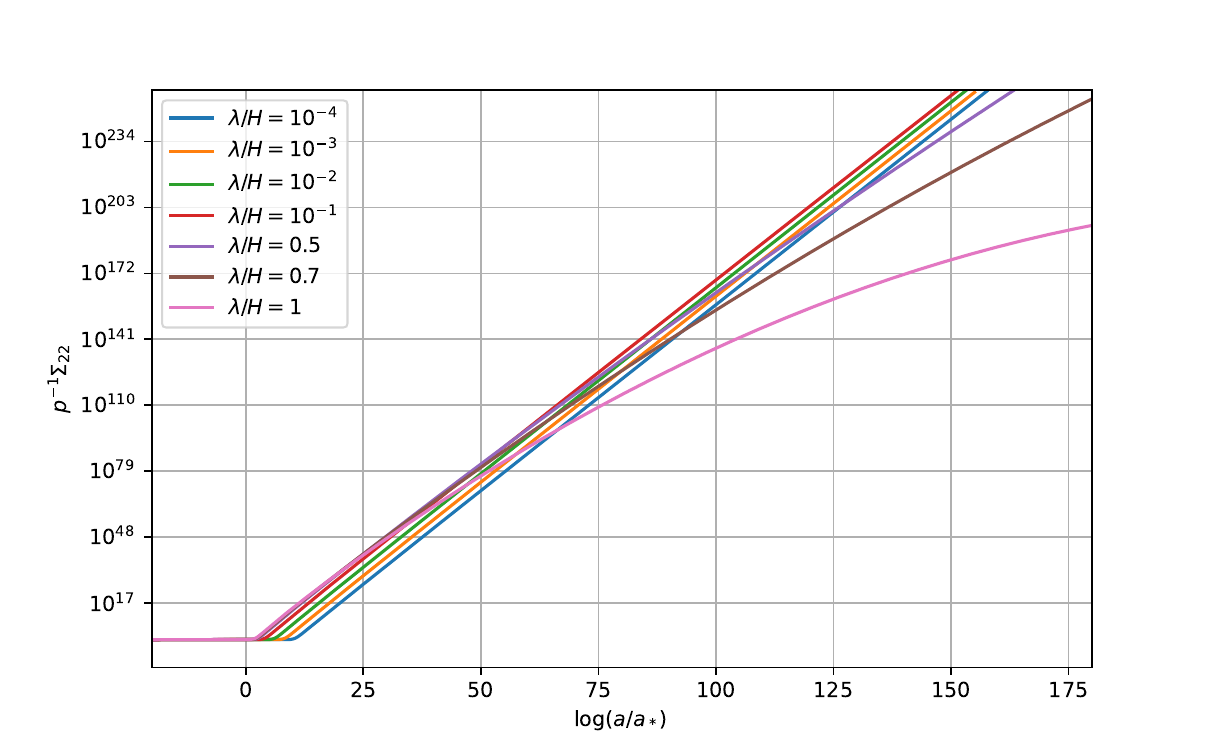}
    \caption{Numerical solution of $p^{-1}\Sigma_{22}$ for different values of $\lambda/H$. }
    \label{fig:S22}
\end{figure}

Figs.~\ref{fig:S11}–\ref{fig:S22} depict the solutions to the transport equations for a range of couplings. In our code, we simulated 20  $e-$folds of evolution in the sub-horizon regime, ensuring that extending this duration yields consistent final results once the mode has long exited the horizon and is in the super-Hubble regime. As before, we have rescaled the quantities -- in this case, the covariance matrix elements -- to make them dimensionless. Additionally, the value of $p$ was chosen based on the same reasoning discussed earlier, and $a/a_* = 1$ is the instant of horizon-crossing.

\newpage

\subsubsection{Non-perturbative resummation and spurious terms}
One aspect of our solutions that might not be immediately apparent is the inherent non-perturbative resummation of any potential secular divergences. This feature has long been recognized as one of the advantages of using master equation techniques for open quantum systems \cite{Chaykov:2022zro, burgess2008decoherence, Burgess:2018sou, Kaplanek:2019dqu,  Brahma:2021mng, Boyanovsky:2015tba, Boyanovsky:2015jen, Burgess:2015ajz, Colas:2022hlq, Colas:2024xjy, Burgess:2024eng}. An analytical scheme which has been used to find resummations for master equations follows a super-horizon approximation which, in effect, calls for dropping all terms proportional to the decaying mode outside the horizon \cite{Boyanovsky:2015tba, Brahma:2021mng}. Specifically, on applying this to this model resulted in an amplitude for the spectrum of the configuration field ($\Sigma_{11}$, for a given ${\bf p}$,) in the approximate form \cite{Boyanovsky:2015tba,Boyanovsky:2015jen}: 
\begin{equation}\label{eq:anl}
    {\cal P}(p;\tau) := \frac{p^3}{2\pi^2} \frac{1}{a^2} \Sigma_{11} (p;\tau) = \frac{H^2}{2\pi^2} \exp{\left[\alpha(p) \ln (-p\tau)\right]} \exp{\left[-\frac{\lambda^2}{12\pi^2 H^2} \ln^2 (-p\tau)\right]} \langle Q_{\vt p} Q_{-\vt p}\rangle (\tau_*)\;,
\end{equation}
where
\begin{equation} \label{eq:alpha}
    \alpha(p) = \frac{2}{3} \frac{M_R^2 (\tau_0)}{H^2} + \frac{\lambda^2}{6\pi^2 H^2} \left[\ln(-p\tau_0) + 1 \right]
\end{equation}
and $\langle Q_{\vt p} Q_{-\vt p}\rangle (\tau_*)$ represents the correlation between the free–theory growing modes, and is given by $1/2$ at horizon exit. Further, $M_R$ represents a renormalized mass, whose origin (in our framework) is the counterterm that subtracts the divergent term in $\Delta_{11}$ (see \eqref{eq:DT11}). The double exponential ensures a finite result since the second exponential –– having a quadratic dependence on the number of $e-$folds $N := \ln (-p\tau)$ –– will dominate at late times.

To compare our results with the above analytical expression, $\alpha(p)$ must be fully determined. The form presented in \eqref{eq:alpha} arises from equating the renormalization scale to the initial time at the beginning of inflation when the density matrix gets prepared ($\tau_0$). The reason behind this choice shall be discussed shortly. Our treatment of the divergence in $\Delta_{11}$ differs in that we chose a counterterm that precisely cancels out the infinity, leaving only the finite parts shown in \eqref{eq:DT11}. Consequently, no term analogous to $\ln(-p\tau_0)$ contributes to our result, leading us to drop it and consider $\alpha(p) = \lambda^2/(6\pi^2H^2)$ for comparison purposes\footnote{In the final expression in \cite{Boyanovsky:2015tba}, to provide numerical estimates, $p$ is chosen such that the mode would have exited the horizon 50 e-folds before the end of inflation, \textit{i.e.,} $\ln(-p\tau_0) \simeq 50$. On the other hand, the renormalization scheme there had eliminated the constant term $\lambda^2/(6\pi^2H^2)$ in the exponent.}. Although our implementation of the renormalization procedure varies slightly from previous results \cite{Boyanovsky:2015jen} --  mainly due to the fact that a numerical, yet exact, computation that we have executed here would have been significantly difficult otherwise -- note that this small differences appear in the term linear in $N$ and should not matter for the actual resummation which follows from the double exponential.

\begin{figure}[h!]
    \centering
    \includegraphics[width=0.9\textwidth]{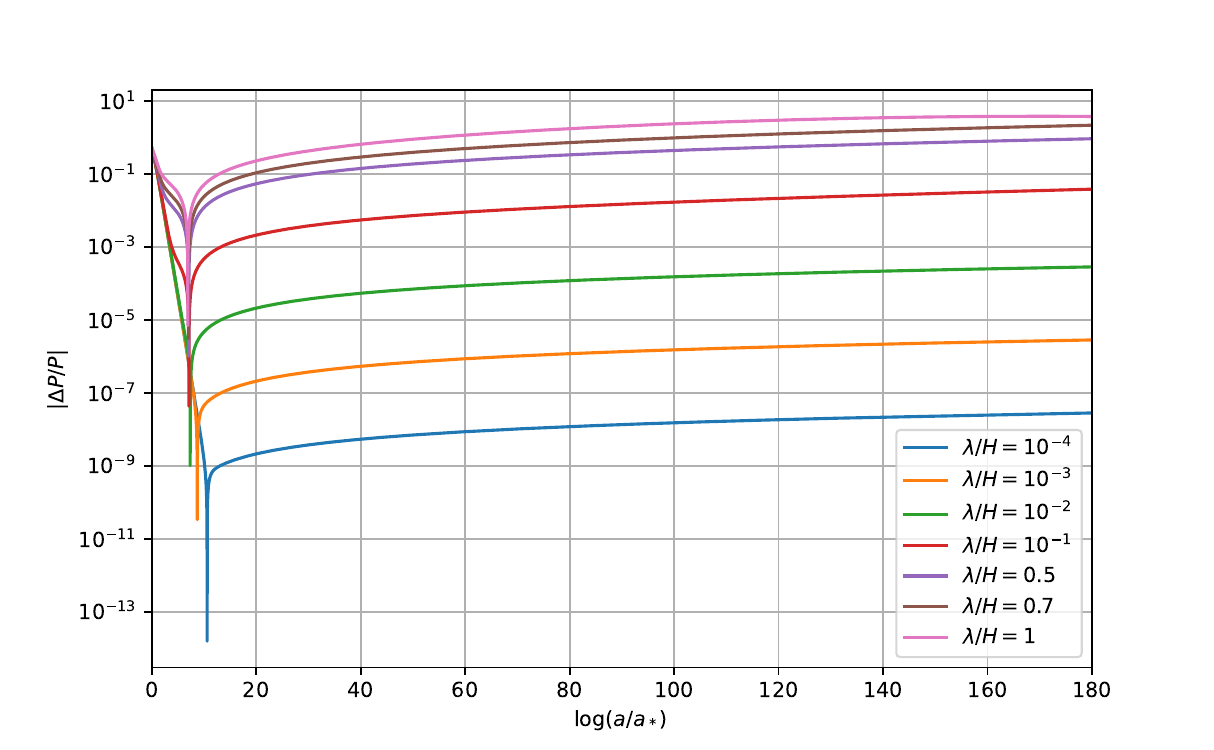}
    \caption{Relative deviation $|\frac{\Delta P}{P}|$ for different values of $\lambda/H$.}
    \label{fig:compa}
\end{figure}

With this caveat in mind, we quantify the consistency of our results with those of \cite{Boyanovsky:2015tba} by defining the relative deviation as:
\begin{equation}\label{eq:Byn}
    \left| \frac{\Delta P}{P}\right| = \left| 1 - \frac{{\cal P}_{\rm SH-approx}}{{\cal P}_{\rm exact}} \right|,
\end{equation}
where ${\cal P}_{\rm SH-approx}$ is given by \eqref{eq:anl}, and ${\cal P}_{\rm exact}$ is the corresponding result obtained with our approach. This function is plotted in Fig.~\ref{fig:compa}, showing a remarkable agreement for small $\lambda/H$, with expected discrepancies for $\lambda \sim H$, the limit in which both the TCL$_2$ approximation and the perturtabative approach in \cite{Boyanovsky:2015tba} break down. To provide further evidence with respect to the consistency of both approaches, we have fitted our results to a double exponential function as described in \eqref{eq:anl}, which ultimately forms an exponential of a quadratic function of $N$. The outcomes of this fitting process are summarized in Table \ref{table1}, focussing on the fit for the quadratic function as it is the most important one for resummation purposes. 

Several insights can be drawn from these fits. Most importantly, the quadratic coefficient in $N$ aligns closely with the prediction in \cite{Boyanovsky:2015tba, Boyanovsky:2015jen}, even for the strongest couplings. As mentioned above, this coefficient is crucial for resummation, indicating that both approaches are comparable in this regard.  The reason behind the discrepancy for the linear term, due to the slightly different ways of implementing the renormalization schemes, has been explained above. Also, it is worth noting that the analytical and numerical approaches predict almost identical coefficients for the term independent of any $e-$foldings ($N^0$). For the former, this value corresponds to the logarithm of the last term in \eqref{eq:anl}, $\ln (1/2) \approx -0.693147$, which aligns well with the fitted functions.

\begin{table}[h!]
\centering
\begin{tabular}{|c|c|c|}
\hline
    $\lambda/H$    &  $\ln \left( \frac{2\pi^2}{H^2} {\cal P}\right)$ & $\propto [\frac{\lambda^2}{12\pi^2 H^2}] N^2$  \\
    \hline 
    $10^{-3}$ & $-8.44291201\times 10^{-9}N^2 + 0.N -0.693147060$ & $0.999938$\\
    $10^{-2}$ & $-8.44384615\times 10^{-7} N^2 + 0.00000005N -0.693135547$ & $1.00005$\\
    $10^{-1}$ & $-8.48517627\times 10^{-5} N^2 + 0.00003749N -0.692702564$ & $1.00494$\\
    $0.5$ & $-0.00210902 N^2 + 0.00009276N -0.663719$ & $0.999129$\\
    $0.7$ & $-0.00412924 N^2 + 0.00006031N -0.63322949$ & $0.998056$\\
    $1$ & $-0.00839741 N^2 -0.00140532 N -0.53644796$ & $0.994549$\\
    \hline
\end{tabular}
\caption{Fitted functions for $\ln {\cal P}$ (rescaled) using the data from Fig. \ref{fig:S11}, from $N=25$ to $N=150$, modeled with a double exponential function as described in \eqref{eq:anl}. The third column displays the ratio between the fitted quadratic coefficient and the analytically predicted value, where a ratio of $1$ indicates perfect agreement.}

\label{table1}
\end{table}

Having shown that the super-horizon approximation that leads to an analytical non-perturbative resummation works quite well when compared to our exact numerical solution, we are now in a position to explain why that is the case and also why does one need an additional approximation in the former case at all. Recall that there was no proper analysis of the spurious terms done in \cite{Boyanovsky:2015tba, Brahma:2021mng}, and yet the results of such a resummation procedure compares quite well to our exact numerical solutions. The reason is the judicious choice of the renormalization scale which was identified with $\tau_0$ in order to minimize the effects from the lower limits of the time integrals involved in the calculations. Although this process is reminiscent to removing spurious terms, it does not quite completely eliminate the contribution of the lower limit of the integral to the master equation. To do so, one needs to implement a further approximation, that of dropping all the decaying modes in the final result. In this way, all contributions of the spurious terms were effectively eliminated in that approach. Although this works fine if one is concerned with computing the value of an observable at the final conformal boundary, the approximation becomes progressively worse at earlier times, when evaluating the evolution of the observable over time as one goes away from the $\tau \rightarrow 0$ limit. On the other hand, in the TCL$_2$ approach, one simply identifies terms which would cancel in a perturbative approach and contain remnants of the initial state, and drops them from the very beginning. In this way, no further approximations are necessary to derive dynamics of cosmological observables. Moreover, apart from not having to impose additional assumptions, the TCL resummation has been shown to better match with exact results for solvable models \cite{Colas:2022hlq} than the super-horizon one.


\subsection{Purity as the determinant of the covariance matrix}
Beside computing the in-in correlation functions, open EFTs are key to studying non-unitary effects such as decoherence (see \cite{Colas:2024xjy,brandenberger1990classical,Calzetta:1995ys,barvinsky1999decoherence,lombardo2005decoherence,lombardo2005influence,martineau2007decoherence,prokopec2007decoherence,burgess2008decoherence,sharman2007decoherence,campo2008decoherence,anastopoulos2013master,Nelson:2016kjm,martin2018non,Oppenheim:2022xjr,DaddiHammou:2022itk,Sharifian:2023jem,Ning:2023ybc,joos2014decoherence,Burgess:2024eng,Bhattacharyya:2024duw,Colas:2024lse,Burgess:2022nwu, Micheli:2023qnc,Ning:2023ybc,Kaplanek:2019dqu,DaddiHammou:2022itk,hollowood2017decoherence,burgess2008decoherence,Chandran:2018wwc,Bhattacharya:2022wpe,Bhattacharya:2023twz,Bhattacharya:2023xvd} for a partial list). In particular, in this paper, we will consider the behaviour of a specific quantum informatic measure, namely purity, for each mode\footnote{As mentioned earlier, for our model different system modes are decoupled from each other at the level of the TCL$_2$ master equation, which greatly simplifies the task at hand.}, which signals the amount of quantum entanglement between the system and the environment. By analysing the purity of the reduced $\rho^\s$, we can study the decoherence (and possibly more exotic phenomena of recoherence or purity-freezing, as a consequence of non-Markovianity \cite{Colas:2022kfu, Colas:2024xjy, Colas:2024ysu}) of the system quantum field.

The purity $\gamma_{\bf p}$ of the system, for a particular momentum mode, is defined as:
\begin{equation}
    \gamma_{\bf p}=\mathrm{Tr}_{\s}\left[(\rho^{\s}_{\bf p})^2\right]\,,
\end{equation}
where $\rho^\s$ is the reduced density matrix, corresponding to the system dofs, as before. When $\gamma_{\bf p}=1$, the system is in a pure state and there is no entanglement between system and environment. When $\gamma_{\bf p}=0$, the system density matrix is maximally mixed.  As a side-note, recall that the purity is indeed independent under field-redifinitions as the trace indicates. From hereon, we will drop the ${\bf p}$ subscript to avoid clutter since the evaluation can be done mode by mode. 

One can simply solve the master equation \eqref{eq:me0} to solve for the reduced density matrix $\rho^\s$ and use that to compute the evolution of the purity. However, an equivalent step would be to use the covariance matrix to do so. More specifically, it is known that for Gaussian states,  we can use the covariance matrix of the system sector   $\mathbf{\Sigma}^\s$\footnote{We introduce the $\s$ superscript here for the covariance matrix to emphasize that this corresponds to the system field alone and is not the full covariance matrix.} to calculate the purity of the system \cite{Ferraro:2005hen}:
\begin{equation}
    \gamma=\frac{1}{\sqrt{4 \mathrm{det}  [\mathbf{\Sigma}^S]}}\,.
\end{equation}
Solving $\rho^\s$ from the master equation by truncating at the TCL$_2$ order is the same as using the covariance matrix to compute the purtiy, while ignoring contributions from higher order correlators, and thus our approximation is self-consistent.

Although we can solve for each of the components of the covariance matrix individually and plug them into the above equation, in practice, it is simpler to derive the equation for the determinant of covariance matrix from \eqref{transporteqn} as
\begin{equation}\label{purity}
    \frac{\mathrm{d}}{\mathrm{d}\tau} \mathrm{det}[\mathbf{\Sigma}^\s] =\mathrm{Tr}[\mathbf{\Sigma}^\s \mathbf{D}] - 4 \mathbf{\Delta}_{12} \ \mathrm{det}[\mathbf{\Sigma}^\s]\,.
\end{equation}
We choose initial conditions such that the $\s$ field is in the Bunch-Davies vacuum. Explicitly, these are given by $\Sigma^\s_{11} = 1/(2p), \Sigma^\s_{22} = p/2, \Sigma^\s_{12} = \Sigma^\s_{21} = 0$, which implies that ${\rm det}[{\bf \Sigma}^\s]=1/4 \Rightarrow \gamma =1$ as $\tau_0 \rightarrow -\infty$. 

\begin{figure}[h]
    \centering
    \includegraphics[width=0.8\textwidth]{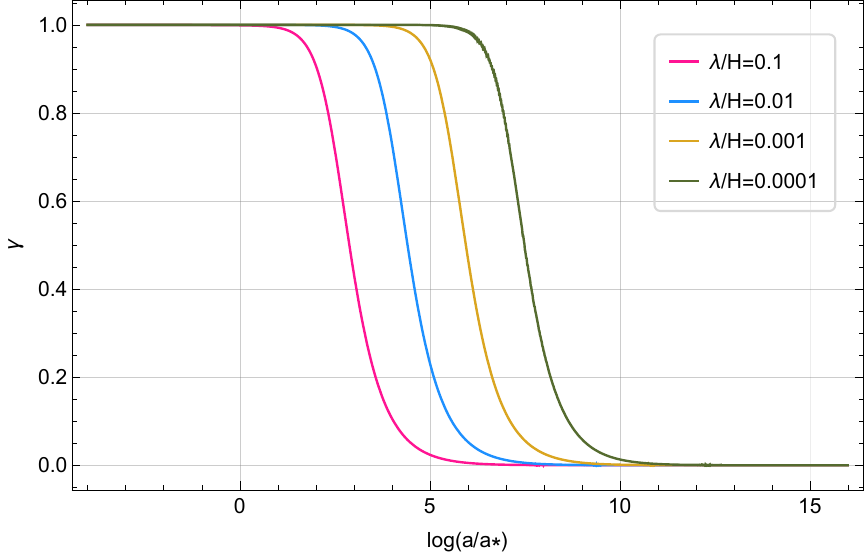}
    \caption{There is rapid decoherence phase occurring right after horizon crossing. As expected, for a system with weaker interaction with environment, the loss in purity occurs at later times.}
    \label{fig:purity}
\end{figure}

With these initial conditions, we solve for the above differential equation for the determinant of ${\bf \Sigma}^\s$ and use that to solve for $\gamma$. For a  mode that spends $10$ e-folds before horizon exit, we solve the above equation and compare the dynamical behavior of purity for different values of $\lambda$ as a ratio over $H$. As shown in Fig. \ref{fig:purity}, the system evolves rapidly to a mixed state due to its coupling with the environment. We conclude that decoherence proceeds very efficiently for this model supporting previous findings \cite{hollowood2017decoherence} derived in a different manner. 

However, this raises an important question. In \cite{Colas:2022kfu, Colas:2024ysu}, it was shown that the purity decreases from 1 and then bounces back up to some value, when considering the (Gaussian) coupling of an adiabatic mode with an entropic one. This was rightly pointed out to be an effect of the non-Markovian nature of the system-environment coupling for an accelerating background. In the previous sub-sections, as was also done in \cite{hollowood2017decoherence} using different approximations, we have demonstrated that the corrections to $\Sigma_{11}$ comes from the non-local part of the memory kernel, and thus it is established that this model is also non-Markovian. Why do we then not see such recoherence or purity-freezing in this case?

This can be understood if we focus on different terms appearing in  \eqref{purity}. The first term on the RHS of that equation is characterized by diffusion, and $D_{11}$ is the dominant contribution to this term due to the initial conditions mentioned above. (Recall that $D_{22}$ is zero for our model and $D_{12}$, which does have a non-local origin, multiplies the off-diagonal terms in ${\bf \Sigma}^\s$.) It can be easily seen that the dynamics is primarily driven by this first term on the RHS which dominates over the second  piece featuring the dissipative $\Delta_{12}$ term (at least, until very late-times once the system classicalizes). Furthermore, the dissipative term forms the homogeneous part of the equation and since $\Delta_{12}$ is negative for this model, the overall sign becomes positive for this term. Consequently, this yields an exponentially growing solution for the determinant, indicative of a decline in purity, \textit{i.e.,} decoherence. However, the main reason why we see no recoherence or purity-freezing is that the diffusion term $D_{11}$, which dominates the evolution of the determinant of ${\bf \Sigma}^\s$, is, in turn, dominated by the local part of the memory kernel as shown in Fig. \ref{fig:lnl} and in \eqref{localvnl}. Consequently, it also indicates that if we were to consider a model in which the coupling for a time-dependent background such that the diffusion terms are primarily derived from the non-local part of the kernel, then we might be able to see such exotic phenomena such as recoherence \cite{Colas:2024xjy}.

\begin{figure}[h]
    \centering
    \includegraphics[width=0.9\textwidth]{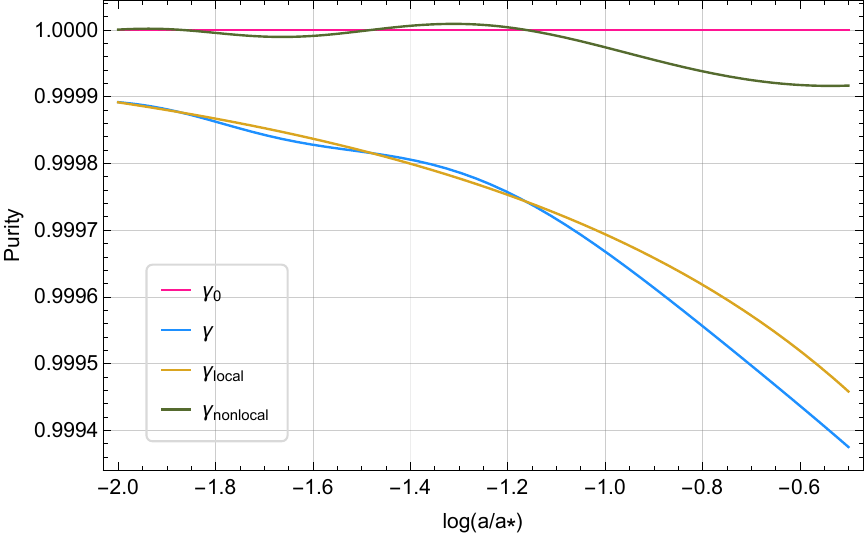}
    \caption{The pink plot, labelled as $\gamma_0$, is the purity for system with no interaction with environment, which is what is expected when the state remains pure. The green plot is the effect coming from non-local part in TCL$_2$ equation, we can see that it leads to oscillations due to information exchange between the system and environment. Comparing the purity when all the terms are retained  (blue plot) with the one when only the local terms are kept (yellow plot), shows that the non-local terms have very little effect on the way purity evolves. This is why the system undergoes decoherence and inevitably evolves to a mixed state once the mode crosses the horizon.}
    \label{fig:puritycompare}
\end{figure}

To drive home this message even more forcefully, let us solve \eqref{purity} separately with only local and non-local components, where we note that while the non-local effects leave an oscillating flow of information between the system and environment, decoherence is purely driven by the local component of $D_{11}$.  We need to zoom in the plot (Fig. \ref{fig:puritycompare}) for the evolution of the purity to see this. We plot sub-horizon evolution of the purity for this since, on artificially only keeping the non-local contributions to the transport equations, the plot for $\gamma_{\text{non-local}}$ diverges after horizon-crossing. Nevertheless, when all the terms are kept (plot corresponding to $\gamma$), the purity is perfectly well-behaved and shows efficient decoherence, and closely follows the plot for purity if we had (again, artificially) dropped all the non-local terms in the transport equations.

For comparison, we also evaluate the purity of the reduced density matrix using perturbation theory (PT). Since $\mathrm{TCL}_{n}$ has been shown to contain all the terms appearing in $\mathrm{PT}_{n}$ and some other higher order $m>n$ terms \cite{Colas:2023wxa}, we get the $\mathrm{PT}_{2}$ equation by terminating the $\mathrm{TCL}_{2}$ equation at $\lambda^2$ order. Accordingly, the perturbative transport equation, at leading order, is given by
\begin{equation}
\frac{\mathrm{d} \boldsymbol{\Sigma}^{(2)}_{\mathrm{PT}}}{\mathrm{d} \tau}=\boldsymbol{\omega} \boldsymbol{H}^{(2)} \boldsymbol{\Sigma}^{(2)}_{\mathrm{PT}}-\boldsymbol{\Sigma}^{(2)}_{\mathrm{PT}} \boldsymbol{H}^{(2)} \boldsymbol{\omega}+\boldsymbol{\omega} \boldsymbol{\Delta} \boldsymbol{\Sigma}^{(0)}-\boldsymbol{\Sigma}^{(0)} \boldsymbol{\Delta} \boldsymbol{\omega}-\boldsymbol{\omega} \boldsymbol{D} \boldsymbol{\omega}-2 \boldsymbol{\Delta}_{12} \boldsymbol{\Sigma}^{(0)}\,.
\end{equation}
The calculation of the perturbative purity shows that $\mathrm{TCL}_{2}$ solution is almost identical to $\mathrm{SPT}_{2}$ solution, which is a signature of validity for $\mathrm{TCL}$ method. It also points to the real advantage of using the TCL formalism lies in its ability to resum secular divergences which appear for some physical observables, such as $\Sigma_{11}$ as shown in the previous sub-section when computed perturbatively, and not so much for purity in this model. 

\begin{figure}[h]
    \centering
    \includegraphics[width=0.9\linewidth]{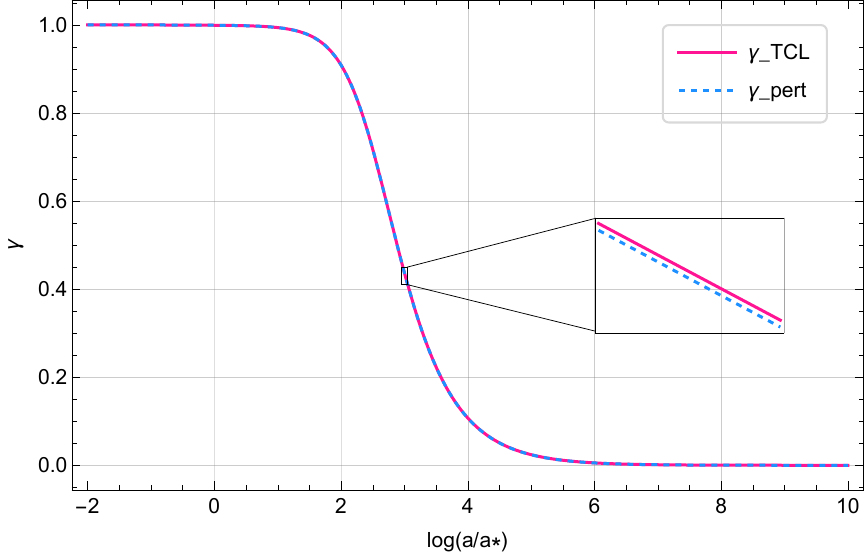}
    \caption{The purity calculated from $\mathrm{TCL}_{2}$ method is very close to the one computed from perturbation theory at second order of $\lambda$. Here we use $\lambda/H=0.1$ case as an example.}
    \label{fig:enter-label}
\end{figure}

\begin{figure}[h]
    \centering
    \includegraphics{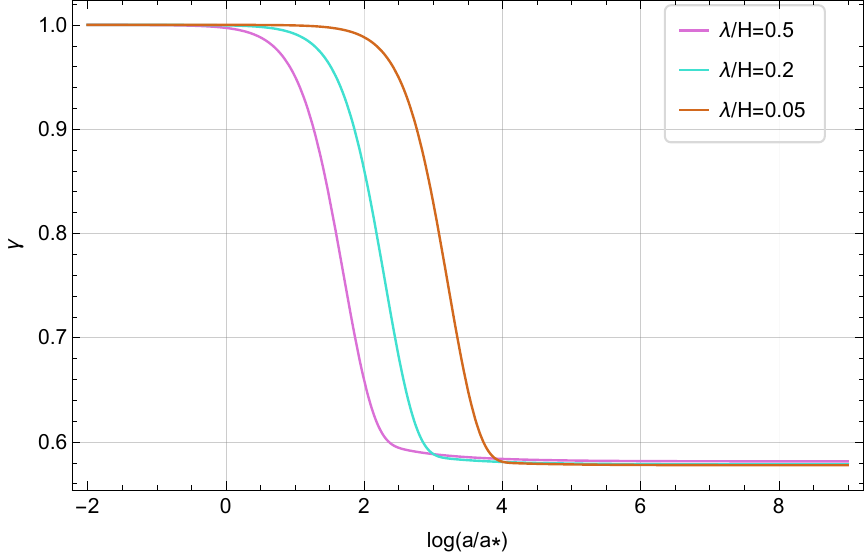}
    \caption{The purity freezes at round $\gamma=0.58$ for different value of $\lambda$'s.}
    \label{fig:spurious_purity}
\end{figure}

Finally, following the expressions of the spurious term $F_i(\tau, \tau_0)$, as computed in the Appendix, we also consider the solution for purity on including these terms. Since the spurious terms dominate at late-times, the purity naturally behaves radically differently in this case. In general, if we do not drop these terms, the system will go through incomplete decoherence and stabilize itself around some non-zero value of purity, as shown in Fig. \ref{fig:spurious_purity}, which is another physical motivation to drop such terms.

\section{Conclusions}
What have we learnt from revisiting this toy model? Firstly, we have shown that secular late-time divergences can appear from the non-local part of the memory kernel and show up in the quantum corrections to the power spectrum of the system field. This is made abundantly clear by the fact that even if we drop the local term in the kernel, the secular divergences in the dissipative terms persist in our model. This is in stark contrast to previous computations for quantum corrections to mode functions of scalar or photon fields due to graviton loops, where the local terms in the evolution equation always dominate at late-times and were responsible for secular divergences. More explicitly, for our model, the solution for the power spectrum is numerical. The leading order expansion of this quantity in the coupling constant $\lambda^2$ is in general what one would have obtained in perturbation theory (in the absence of any resummations). Looking up Table \ref{table1}, the leading order expansion can be written (for, say, $\lambda/H=10^{-3}$),  as $\mathcal{P} ^{\mathrm{PT}}\propto e^{-6.93147060\times 10^{-1}}(1-8.44291201\times 10^{-9}N^2)$, which clearly denotes secularly divergent terms, the superscript `PT' reminding us that this would be the result in perturbation theory\footnote{This, of course, can also be demonstrated from the results of \cite{Boyanovsky:2015tba} 
\begin{equation}
    {\cal P}^{\mathrm{PT}}(p;\tau)= \frac{H^2}{2 \pi^2}\langle Q_{\vt p} Q_{-\vt p}\rangle (\tau_*)\left\{1-\frac{\lambda^2}{12\pi^2 H^2}\left[2 \mathrm{ln}(-p\tau)-\mathrm{ln}^{2}(-p\tau)\right] \right\}\,.
\end{equation}}.
Thus, quantum corrections to a physical observable can indeed have secularly growing contributions from non-Markovian terms and, what is more,  the time-convolutionless form of the master equation is nevertheless able to dynamically resum such terms and render finite results. Although some of these observations have been made in the past, we correct some misconceptions regarding the Markovian approximation, namely that having a time-local master equation does not necessarily imply that the system must be Markovian.

This gives us a segue into how we have also shed light on an important aspect of such resummations which have been ignored in the past -- spurious terms which arise from the lower (time) limit of the kernel integration as remnants of the initial state. This can be very cleanly handled using the TCL formalism and we have shown why such terms must be dropped by hand in order to have a meaningful resummation of physical observables. It makes sense to drop such terms since they are cancelled, between themselves, in a perturbative treatment to any order. Furthermore, we have resolved the interesting paradox of why previous computations of the power spectrum \cite{Boyanovsky:2015jen, Boyanovsky:2015tba} yielded accurate results even when such a treatment of spurious terms were lacking. This was due to a judicious choice of the renormalization scale coupled with an additional \textit{ad hoc} approximation of dropping the decaying mode contributions in the final answer. Although this approximation is good one for computing the power spectrum at the conformal boundary, it falls short in tracking the evolution of the elements of the covariance matrix when going to earlier times. The TCL formalism does not demand such approximations and render them redundant by applying a consistent analysis of the spurious terms. Another reason to have more faith in the TCL$_2$ resummation over the super-horizon one employed in \cite{Brahma:2021mng,Boyanovsky:2015tba} is that for exactly solvable Gaussian models, the former has been shown to have a closer fit with exact results \cite{Colas:2022hlq}. 

Secondly, what our model shows is that although the memory kernel has a non-local part, the diffusion term which controls decoherence is dominated by the local term. In this context, we clarify that unlike previous claims, there is no emergent Markovian behaviour after horizon-crossing. First, this is evidenced by the fact that the corrections to the power spectrum, coming from non-local part of the kernel, survive in the $\tau \rightarrow 0$ limit. Just because the system density matrix attains ``positivity'' at late-times  \cite{hollowood2017decoherence}, which is a quantitative measure of classicality of the system, does not mean that system has become Markovian. In fact, as has been shown in \cite{hollowood2017decoherence} itself, one of the eigenvalues of the dissipator matrix remains negative at all times, thus clearly satisfying the definition of a non-Markovian system. Rather, what happens is that $D_{11}$ is dominated throughout  by the contributions coming to it from local terms, which get rather pronounced at after horizon-crossing and at late times. Although $D_{11}$ plays the crucial role for decoherence, its contribution to the quantum correction to the power spectrum is negligible and thus different physical observables can depend on the local and non-local pieces of the memory kernel. One of the main reasons for the belief that late-time secular growth in inflationary models typically come from local terms is stochastic inflation since the latter invokes a white noise for modelling the quantum diffusion due to the sub-horizon modes. Our work shows that there might be a case for considering the non-local corrections to the noise term for stochastic inflation as well. Indeed, in this model, it is the dissipation terms that appear solely from the non-local terms and hence it will be interesting to consider a model in the future where the diffusion (or the noise) terms depend exclusively on the non-Markovian piece of the memory kernel. We leave constructing such a realistic model for future work \cite{US:non-local_D22}. Our results also find that purity is not a sufficiently useful tool to probe the Markovianity of the model. In \cite{Colas:2022kfu,Colas:2024ysu}, it was shown that a non-Markovian kernel can lead to recoherence. What we find is that this is, however, not a smoking gun signal of non-Markovianity.

To the best of our knowledge, this is the first application the TCL formalism to a cosmological model involving a non-linear interaction generalizing beyond strictly Gaussian models. Our results, although sometimes technical, would serve well as benchmarks against applying this particular master equation formalism to future realistic models for inflation.

\section*{Appendix: Spurious Terms}
As mentioned in Sec-3, the lower limits of integrals \eqref{d11i}-\eqref{dt12i} give rise to so-called spurious terms. These terms, initially studied in the context of the cosmological Caldeira-Leggett model \cite{caldeira1981influence,banerjee2023thermalization}, were found to dominate at late times for all coefficients, which is unexpected for time-local (or quasi-time-local) master equations. In the same context, these terms were shown to be absent for perturbative solutions to any given order, as explained earlier. If these spurious terms are considered, they could lead to a breakdown of the resummation of the power spectrum, contradicting the exact results available for such a model.

Clearly, it is crucial to check the behaviour of these terms within the model we are working with. This not only completes the TCL$_2$ description, but also adds this is necessary when considering non-Gaussian models such as this one, for which exact results are not available. An expectation that was articulated in \cite{Colas:2023wxa} was that for more realistic interactions, going beyond the Gaussian case, it might be true that such spurious terms (coming from the lower limit of the integrals) would get suppressed. We want to show that this is certainly not the case for our model, and this has consequences for the non-perturbative resummation as has been explained in Sec-4.

\begin{figure}[h!]
    \centering
    \includegraphics[width=0.9\textwidth]{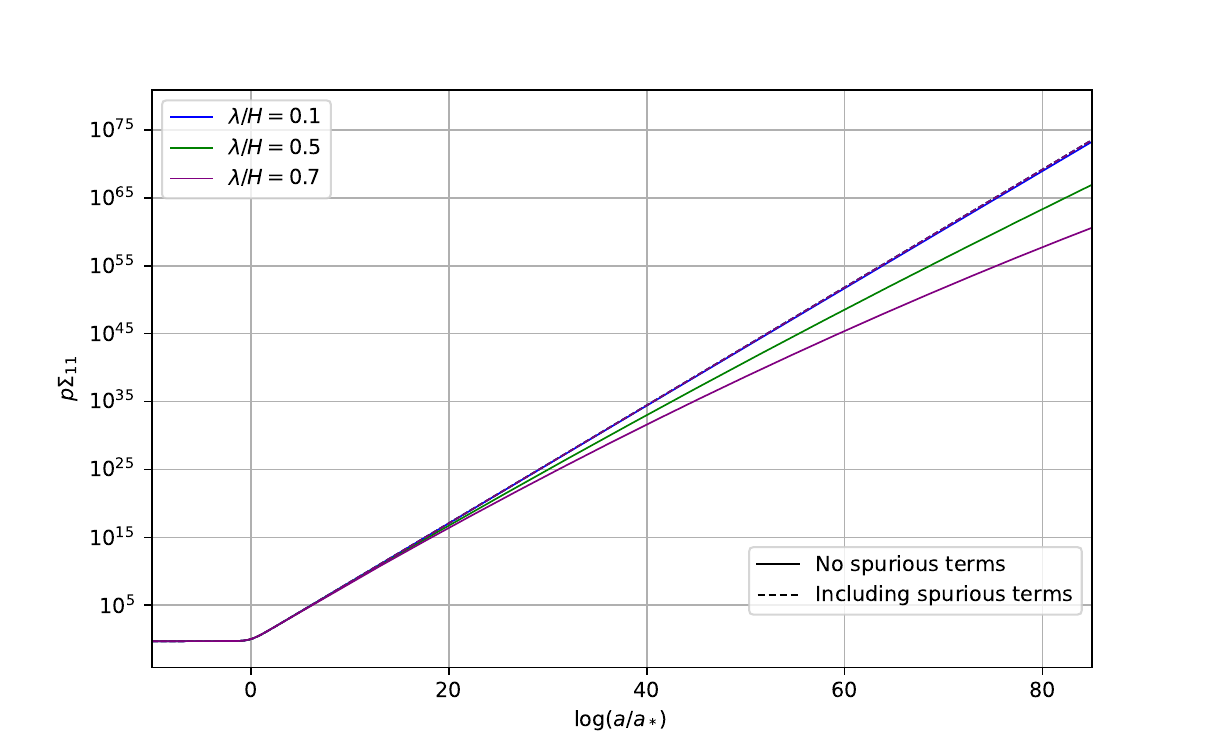}
    \caption{$p\Sigma_{11}$ for different values of $\lambda/H$, considering spurious terms (dashed lines) and without considering them (solid lines). }
    \label{fig:sp}
\end{figure}

\begin{figure}[h!]
    \centering
    \includegraphics[width=0.9\textwidth]{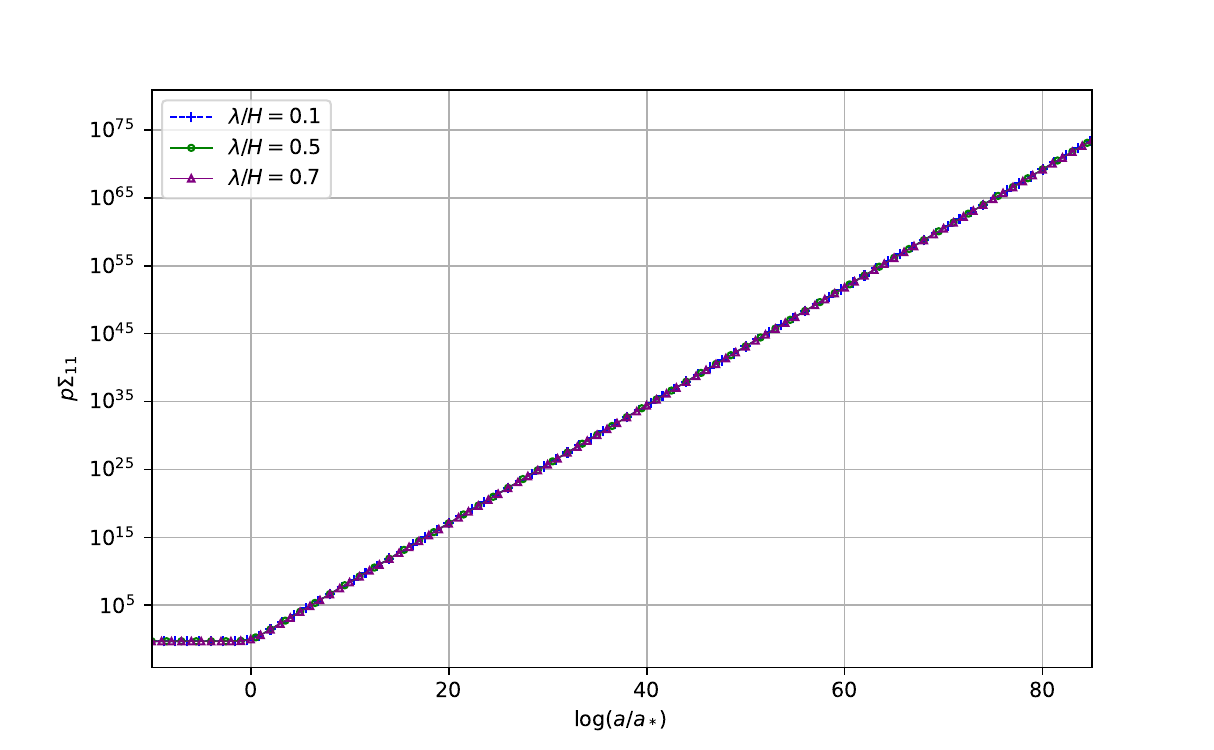}
    \caption{$p\Sigma_{11}$ for different values of $\lambda/H$, considering spurious terms.}
    \label{fig:sp2}
\end{figure}

For the reader's convenience, we show again the explicit form of the master equation coefficients (keeping our color code for the local and divergent terms):
\begin{align}
    D_{11} & = -\frac{\lambda^2}{8\pi^2 H^2 p \tau^3} \Big[ \gamma_E + \ln(-2p\tau) - {\rm Ci}(-2p\tau) \left[ \cos(2p\tau) + p \tau \sin(2p\tau) \right] \nonumber \\
    & + {\rm Si} (2p\tau) \left[ p\tau \cos(2p\tau) - \sin(2p\tau) \right] \Big] + \textcolor{blue}{\frac{\lambda^2}{4\pi H^2 \tau^2}} + F_1[\tau,\tau_0]\;,
\end{align}
\begin{align}
    \Delta_{11} & = \frac{\lambda^2}{8\pi^2 H^2 p \tau^3} \Big[ p\tau \left[ \gamma_E -\ln(-2 p\tau) \right] + {\rm Ci}(-2p\tau) \left[p\tau \cos(2p\tau) - \sin(2p\tau) \right] \nonumber \\
    & + {\rm Si}(2p\tau) \left[ \cos(2p\tau) + p \tau \sin(2p\tau) \right]
 \Big] + \frac{\lambda^2}{4\pi^2 H^2 \tau^2} \textcolor{red}{\ln(2p\epsilon)} + F_2 [\tau,\tau_0]\;,
\end{align}
\begin{align}
    D_{12} & = \frac{\lambda^2}{16 \pi^2 H^2 p^3 \tau^4} \Big[ (1+p^2 \tau^2) \left[ \gamma_E + \ln(-2p\tau) \right] + {\rm Ci}(-2p\tau) \left[ (-1 + p^2 \tau^2) \cos(2p\tau) - 2p\tau \sin(2p\tau) \right] \nonumber \\
    & + {\rm Si}(2p\tau) \left[ 2p\tau \cos(2p\tau) + (-1 + p^2 \tau^2) \sin(2p\tau) \right]\Big] + \textcolor{blue}{0} + F_3[\tau,\tau_0]\;,
\end{align}
\begin{align}
    \Delta_{12} & = \frac{\lambda^2}{16\pi^2 H^2 p^3 \tau^4} \Big[ 2 p \tau + {\rm Ci}(-2p\tau) \left[ \sin(2p\tau) - p\tau (2 \cos(2p\tau) + p\tau \sin(2p\tau)) \right] \nonumber \\
    & + {\rm Si}(2p\tau) \left[ (-1+p^2 \tau^2) \cos(2p\tau) - 2p\tau \sin(2p\tau) \right] \Big] + F_4 [\tau,\tau_0]\;,
\end{align}
where $F_i$ are the expressions coming from the lower limit of the integrals. \sout{As for the upper limits,} Analytical expressions are available for these functions, namely
    \begin{align}
        F_1 (\tau, \tau_0) & \approx \frac{\lambda^2}{16\pi H^2 p \tau^3} \big[ \sin (2p\tau) - p\tau (2 + \cos(2p\tau)) \big] +  {\cal O} (\tau_0^{-1})\;, \\
        F_2 (\tau, \tau_0) & \approx \frac{\lambda^2}{16\pi H^2 p \tau^3} \big[ -2 + \cos(2p\tau) + p\tau \sin(2p\tau) \big] + {\cal O} (\tau_0^{-1})\;, \\
        F_3 (\tau,\tau_0) & \approx \frac{\lambda^2}{16\pi H^2 p^3 \tau^4} \big[ p \tau \cos(p\tau) - \sin(p\tau) \big] \big[ \cos(p\tau) + p \tau \sin(p\tau) \big] + {\cal O} (\tau_0^{-1})\;, \\
        F_4 (\tau, \tau_0) & \approx \frac{\lambda^2}{32 \pi H^2 p^3 \tau^4} \big[ 2(1+ p^2 \tau^2) - (1-p^2 \tau^2) \cos(2p\tau) - 2 p \tau \sin(2p\tau)  \big] + {\cal O} (\tau_0^{-1})\;,
    \end{align}
where we have isolated the nonzero parts in the limit $\tau_0 \rightarrow -\infty$. Notice how even after doing this, there is a surviving contribution at late times, which we should now check to see if they are dominant over the others. Expanding these functions in the limit $-p\tau \ll 1$, we get:
\begin{equation}
     F_1 (\tau)  \approx - \frac{\lambda^2}{16\pi H^2 \tau^2}\;, \quad 
    F_2 (\tau)  \approx - \frac{\lambda^2}{16\pi H^2 p \tau^3}\;, \quad
    F_3 (\tau)  \approx -\frac{\lambda^2}{48\pi H^2 \tau}\;, \quad
    F_4 (\tau) \approx \frac{\lambda^2}{32\pi H^2 p^3 \tau^4}\;.
\end{equation}
Compare this to the non-spurious late-time behaviour of the master equation coefficients, given by
\begin{align}
    D_{11} & \approx \frac{\lambda^2 p}{8\pi^2 H^2 \tau} + \frac{\lambda^2}{4\pi H^2 \tau^2}\;, \\
    \Delta_{11} & \approx \frac{\lambda^2}{4\pi^2 H^2 \tau^2} [1 - \ln(-2p\tau) + \ln (2p\epsilon)]\;, \\
    D_{12} & \approx \frac{\lambda^2}{16\pi^2 H^2 p\tau^2}\;,\\
    \Delta_{12} & \approx \frac{\lambda^2}{72\pi^2 H^2 \tau} [-7 + 3\gamma_E + 3 \ln(-2p\tau)]\;.
\end{align}
Notice how the spurious terms are at least comparable to these expressions, with only the spurious part of $D_{12}$ being subdominant. Fig. \ref{fig:sp} presents the solution of the transport equation for $\Sigma_{11}$, comparing cases with and without the inclusion of spurious terms. It is immediately noticeable that when spurious terms are excluded, the solutions increase at a lower rate, a signal of resummation. Further, for the time range considered, the solutions obtained by including spurious terms look insensitive to the choice of $\lambda$, as illustrated in Fig. \ref{fig:sp2}.

\section*{Acknowledgements} 
SB thanks Drazen Glavan for interesting discussions regarding secular divergences from non-local terms that spurred this work. The authors also thank Thomas Colas for numerous discussions about open quantum systems in cosmology.

SB is supported in part by the Higgs Fellowship and by the UK Science and Technology Facilities Council (STFC) Consolidated Grant “Particle Physics at the Higgs Centre”. JCF is supported by the STFC under grant ST/X001040/1.
XL is supported in part by the Program of China Scholarship Council (Grant No. 202208170014).

For the purpose of open access, the author has applied a Creative Commons Attribution (CC BY) licence to any Author Accepted Manuscript version arising from this submission.

\printbibliography

\end{document}